\documentclass[mlmain]{jmlr}

\usepackage{longtable}% for long tables
\usepackage{threeparttable}% for aligning table footnotes

\usepackage{booktabs}
\usepackage[load-configurations=version-1]{siunitx} % newer version
\jmlrvolume{}
\firstpageno{1}
\editors{List of editors' names}

\jmlryear{2026}
\jmlrworkshop{Symmetry and Geometry in Neural Representations}

\title[Geometric Assumptions in Active Learning]{Benchmarking Geometric Assumptions in Active Learning for Non-Normal Neural Dynamics}

 \author{\Name{Melina M{\"u}ller} \Email{m.m.mueller@sms.ed.ac.uk}\\
    \Name{Matthias H. Hennig} \Email{m.hennig@ed.ac.uk}\\
    \addr School of Informatics, University of Edinburgh, Edinburgh, United Kingdom\\    
    \AND    
    \Name{Luke Bashford} \Email{luke.bashford@newcastle.ac.uk}\\
    \addr Faculty of Medical Sciences, Newcastle University, Newcastle, United Kingdom\\    
    \AND    
    \Name{Nina Kudryashova} \Email{nkudryas@ed.ac.uk}\\
    \addr School of Informatics, University of Edinburgh, Edinburgh, United Kingdom\\
}

\begin{document}

\maketitle

\begin{abstract}
  As systems neuroscience shifts toward active, closed-loop control, efficient system identification within limited experimental sessions is critical.
  However, emerging active learning (AL) strategies lack a standardized framework to compare their differing geometric assumptions and objectives.
  To address this, we introduce a common formalism for A-optimal active learning of linear neural dynamics, alongside a comprehensive benchmark designed to evaluate active learning approaches for neural population dynamics.
  We systematically evaluate state-of-the-art sampling methodologies under challenging, biologically relevant conditions: low intrinsic dimensionality and non-normal dynamics.
  Informed by this evaluation, we propose two novel active learning methods: one combining the best modeling choices of the existing approaches (`Hybrid'), and another one specifically exploiting transient non-normal amplification (`Balanced'). Both methods achieve significantly faster convergence to latent dynamics, particularly in the non-normal case, compared to existing approaches.
  Ultimately, this benchmark provides critical insights into the assumptions and limitations of current active learning approaches, synthesizing them into a robust approach suitable for identifying computations performed by biological neural networks.
\end{abstract}

\section{Introduction}
Advances in recording technologies have led to an increase in the number of simultaneously recorded and stimulated neurons. This access enables a shift from passive correlation analyses of neural activity to inferring their causal relationships through targeted perturbations. Not only does this allow for more mechanistic insights into neural circuit connectivity, but also offers the opportunity to control neural states. However, choosing the optimal stimulation input to identify the neural system governing the recorded neural dynamics remains a challenge. With increased access, exhaustively testing all possible stimulation parameters combinations has become infeasible even across multiple recording sessions. Randomly selecting subsets of perturbations is therefore a common solution \citep{o2022direct, bolus_state-space_2021}, but might waste resources on uninformative stimulations. Instead, more targeted approaches are needed for sub-selecting the most informative perturbations.

Some recent approaches address this by inferring local neural dynamics from responses to targeted optical or electrophysiological microstimulation. These either sub-select optimal stimulations before the start of a session~\citep{tentori2025spontaneous, nejatbakhsh2023predicting, marshel2019cortical}, target perturbations using state-prediction heuristics~\citep{minai_miso_nodate}, or adaptively choose the best next parameters at each step based on the stimulus-response data gathered so far.
For the latter, two such active learning (AL) methods have been proposed so far by \citet{wagenmaker_active_nodate} and \citet{ogino_designing_2025}.
Both approaches use principles from Optimal Experimental Design for system identification by mathematically quantifying how the perturbation input can be chosen to reduce the error of the fitted system parameters.
In particular, they both derive an optimisation term based on A-optimality (Average optimality), which aims to minimise the average variance of the parameter estimates~\citep{pukelsheim2006optimal}.
However, their exact goals and assumptions differ, with \citet{ogino_designing_2025} aiming to capture the fastest vanishing modes in the full-rank neural system, while \citet{wagenmaker_active_nodate} concentrates on identifying low-rank dynamics.

The hypothesis that neural activity evolves along low-dimensional manifolds has emerged as a dominant governing principle of neural dynamics \citep{gallego2018cortical,perich2025neural}. However, the evaluation of system identification by~\citet{wagenmaker_active_nodate} was limited to generic linear systems, without biologically-relevant constraints on connectivity.
Neural population dynamics are inherently non-normal due to biological constraints like Dale's law~\citep{murphy2009balanced, hennequin2012non} and circuit motifs such as feedforward chains~\citep{goldman2009memory}, both of which manifest in transient signal amplification.
It was recently shown that non-normal systems can easily be misidentified, i.e. a non-normal feedforward chain can be mistaken for a normal line attractor~\citep{qian2024partial}.
Yet, to the best of our knowledge, no targeted sampling study has explicitly considered the non-normality of neural population dynamics.
It is therefore important to establish whether current AL methods are effective for both normal and non-normal systems.

The main goal of this work is to provide a common formalism for A-optimal AL of linear neural dynamics.
We focus on systems with non-trivial geometry shaped by biological constraints, such as non-orthogonal eigenmodes (non-normal systems) and low-dimensional manifold structure (low-rank connectivity).
We synthesized the different assumptions of these A-optimal AL methods into a single mathematical framework, and tested their combinations on low-rank dynamical systems with varying non-normality.
We find that different learning goals lead to a trade-off in performance across measures, such as identifying a latent versus the full system. Finally, we propose a new method that unifies several ideas across these approaches and is uniquely capable to correctly identify non-normal dynamics.

\section{Dimensions of active learning optimisation}
\subsection{Analytical comparison}
Neural population dynamics can be modelled as a linear dynamical system
\begin{equation}
  \mathbf{x}(t+1) = \mathbf{W} \mathbf{x}(t) + \mathbf{B} \mathbf{u}(t)
  \label{Eq:discrete_linear_sys}
\end{equation}
in which the recurrent weights $\mathbf{W}$ are particularly insightful for characterizing the computation performed by the neural population~\citep{vyas2020computation}.
To minimize the estimation error of these model parameters $\mathbf{M}$ (e.g. $\mathbf{M}=\mathbf{W}$), one can design an experiment that probes the system and optimally chooses the next stimulation input $\mathbf{u}$.
This can be achieved by minimizing the trace of the inverse of the linear Fisher information matrix $\mathcal{I}(\mathbf{M})$ (an A-optimality objective w.r.t. parameters $\textbf{M}$):
\begin{equation}
  \mathbf{u}_\mathrm{A-opt} \stackrel{\text{def}}{=} \arg\min \text{Tr}((\mathcal{I}(\mathbf{M}))^{-1})
  \label{Eq:optimisation_A_opt}
\end{equation}
Two main methods for AL of neural dynamics were recently proposed by~\citet{wagenmaker_active_nodate} and~\citet{ogino_designing_2025}, both following this A-optimality principle. Yet, the assumptions and goals of their approaches differ along various aspects (see Fig.~\ref{fig: fig1_setup}a).

First, $\mathbf{M}=\mathbf{W}$ for \citet{ogino_designing_2025} and $\mathbf{M}=\mathbf{H}$ for \citet{wagenmaker_active_nodate}, where $\textbf{H}$ is the transference matrix mapping the input at $t=0$ to the total elicited neural activity:
\begin{equation}
  \sum_{t=0}^{T} \mathbf{x}(t) = \mathbf{H} \mathbf{u}(0), \quad \text{i.e. }  \mathbf{H}= \sum_{t=0}^{T-1} \mathbf{W}^{t}\mathbf{B}.
  \label{Eq:H_matrix}
\end{equation}
To infer their respective causal connectivity matrix $\mathbf{M}$ both methods derive approximate error terms, which we unify in the following equation:
\begin{equation}
  \mathbf{u} = \arg\min \text{Tr}\left( (\mathbf{P} \boldsymbol{\Sigma}_U \mathbf{P}^\top)^{-1} \right).
  \label{Eq:optimisation_al_error}
\end{equation}
For \citet{ogino_designing_2025} (eq. (8)) $\mathbf{P}=\mathbf{W}$, and for \citet{wagenmaker_active_nodate} (Theorem 1) $\mathbf{P}=\mathbf{V_r}$ (right singular vectors) of $\mathbf{H}$. Note, that neither of these two versions of the error terms (\ref{Eq:optimisation_al_error}) is strictly A-optimal (\ref{Eq:optimisation_A_opt}), not w.r.t. $\mathbf{W}$ nor w.r.t. $\mathbf{H}$.

Second, the methods differ in assumptions on the latent dimensionality. \citet{wagenmaker_active_nodate} demonstrated the efficacy of a low-rank neural network dynamics model. Low-rank assumption constrains the network dynamics to a lower-dimensional linear manifold, which was shown to be a valid assumption, particularly for motor cortical areas~\citep{gallego2018cortical,perich2025neural}. The low-rank constraint is imposed by a truncated singular value decomposition (SVD) $\mathbf{M}=\mathbf{V}_l\mathbf{S}\mathbf{V}_r$, where $\mathbf{V}_l$ and $\mathbf{V}_r$ are the left and right singular vectors, and $\mathbf{S}$ are the diagonalised singular values.
In contrast, \citet{ogino_designing_2025} considers the full-rank dynamics of $\mathbf{W}$, which is a relevant assumption for whole-brain recordings that do not necessarily exhibit lower-dimensional dynamics (Fig.~\ref{fig: fig1_setup}a.ii).
However, they interpret their method by considering how activity transfers along the eigenvectors of the system. This aspect could be considered by instead decomposing $\mathbf{M}$ using eigenvector decomposition (EVD) $\mathbf{M}=\mathbf{Q}_r\mathbf{E}\mathbf{Q}_l$ ($\mathbf{Q}_r$ are the right eigenvectors, $\mathbf{Q}_l$ are the left eigenvectors, and $\mathbf{E}$ are the diagonalised eigenvalues, Fig.~\ref{fig: fig1_setup}a.iii).
Lastly, \citet{wagenmaker_active_nodate} then uses the SVD of $\mathbf{M}$ to optimise the spread of the inputs along the input manifold, i.e. $\mathbf{P}=\mathbf{V}_r$, while \citet{ogino_designing_2025} aims to have elicit isotropic neural activity with $\mathbf{P}=\mathbf{M}$ (Fig.~\ref{fig: fig1_setup}a.iv).

In summary, the method by \citet{wagenmaker_active_nodate} is based on the assumption that neural activity evolves along a low-rank manifold. Minimizing estimation error for a low-rank causal connectivity matrix, derived based on the total activity arising post stimulation, leads to isotropic stimulation along the on-manifold dimensions. In contrast, \citet{ogino_designing_2025} were motivated by whole-brain recordings with full-rank dynamics. Minimizing estimation error in this case leads to enhanced stimulation of the faster decaying terms by ensuring that the evoked activity (rather than input) is distributed isotropically.
Thus the axes of differences (Fig.~\ref{fig: fig1_setup}a) that arise between these two methods are the 1) causal connectivity matrix ($\mathbf{W}$ vs $\mathbf{H}$), 2) system rank (full vs manifold), 3) system modes (EVD vs SVD), and 4) Isotropic space (activity vs input). These difference aspects can be combined in novel ways to up to 16 ($=2^4$) methods. Hence the question is what is the best combination of these ideas, in particular for identifying low-rank non-normal systems.

\subsection{Benchmarking Platform}
\begin{figure}[t]
  \centering
  \includegraphics[width=0.75\linewidth]{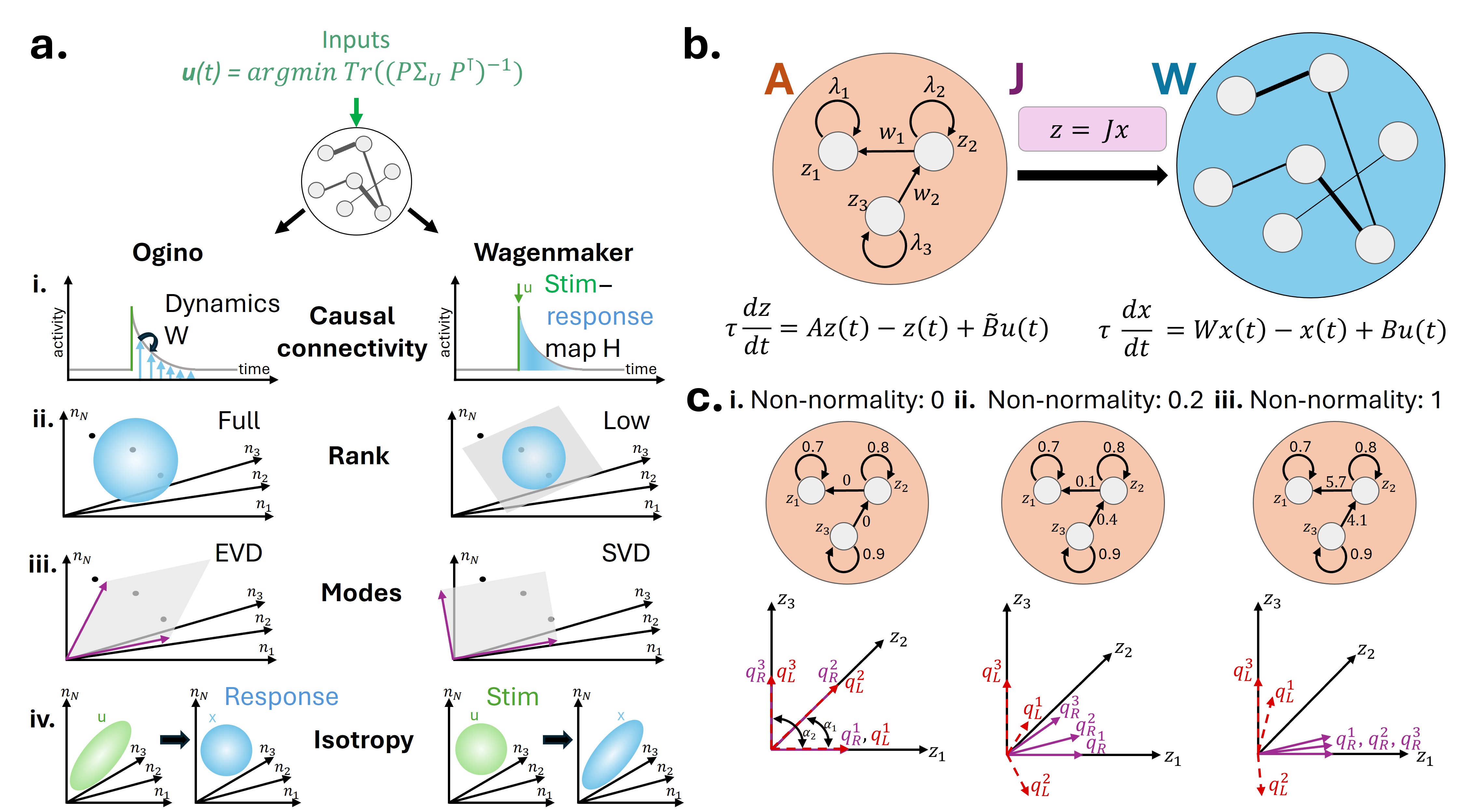}
  \caption{Our benchmark identifies how geometric modeling choices (\textbf{a}) affect dynamical system identification, validated on synthetic data with low-rank dynamics (\textbf{b}) and varying non-normality (\textbf{c}). \textbf{a.} Overview of model assumptions and identification objectives that influence optimal input choice (\ref{Eq:optimisation_al_error}).
    \textbf{b.} Low-rank simulation set-up with feedforward dynamics $\mathbf{z}$, determined by recurrent matrix $\mathbf{A}$, which are projected by $\mathbf{J}$ into neural activity space $\mathbf{x}$ with recurrent matrix $\mathbf{W}$.
    \textbf{c.} Adjusting the feedforward weights $w_1$ and $w_2$, sets the angles $\alpha_1, \alpha_2$ between eigenvectors of $\mathbf{A}$ and thus its non-normality, while keeping the same eigenvalues $\lambda_1,\lambda_2, \lambda_3$. Non-normality leads to the left $\mathbf{q}_l$ and right $\mathbf{q}_r$ eigenvectors not being orthogonal. }
  \label{fig: fig1_setup}
\end{figure}

To evaluate how these modelling choices (Fig.~\ref{fig: fig1_setup}a) affect AL performance in identifying neural network dynamics, we generated synthetic datasets using linear latent dynamics models (Fig.~\ref{fig: fig1_setup}b):
\begin{equation}
  \tau \frac{d\mathbf{z}}{dt} = \mathbf{A} \mathbf{z}(t) - \mathbf{z}(t) + \tilde{\mathbf{B}}\mathbf{u}(t)  + \tilde{\mathbf{w}}(t),
  \label{Eq:latent_lds}
\end{equation}
where $\mathbf{z}(t)$ is the latent state, and matrix $\mathbf{A}$ is a three-dimensional latent neural dynamics matrix we want to identify. Its non-normality strength was defined as the normalised term $\| \mathbf{A}\mathbf{A}^\top - \mathbf{A}^\top\mathbf{A} \| / \| \mathbf{A}\mathbf{A}^\top \|$, which is proportional to maximum transient amplification~\citep{bondanelliCodingTransientTrajectories2020} for most cases. This latent state $\mathbf{z}$ was then embedded into the neural space of 100~neurons $\mathbf{x}$ using a random, semi-orthonormal matrix $\mathbf{J}$ ($\mathbf{z}=\mathbf{J} \mathbf{x}$), leading to the full neural system with recurrent connectivity matrix $\mathbf{W} = \mathbf{J}^\top \mathbf{A} \mathbf{J}$. $\mathbf{W}$ has the same non-normality and top three eigenvectors and eigenvalues as $\mathbf{A}$. Hence, the activity on the neural level $\mathbf{x}(t)$ is:
\begin{equation}
  \tau \frac{d\mathbf{x}}{dt} = \mathbf{W} \mathbf{x}(t) - \mathbf{x}(t) +  \mathbf{B}\mathbf{u}(t) + \mathbf{w}(t),
  \label{Eq:full_lds}
\end{equation}
where $\mathbf{B}$ is the identity matrix, $\tilde{\mathbf{B}}=\mathbf{J} \mathbf{B}$, and $\tilde{\mathbf{w}}=\mathbf{J} \mathbf{w}$ with $ \mathbf{w}\sim \mathcal{N}(\mathbf{0}, \mathbf{\Sigma_d})$. The readout activity is then
$\mathbf{y}(t) = \mathbf{C}\mathbf{x(t)}  + \mathbf{v}(t)$, with $ \mathbf{v}\sim \mathcal{N}(\mathbf{0}, \mathbf{\Sigma_r})$. $\mathbf{C}=\mathbf{I}$ assumes full observability.

For our benchmark, we created three such linear dynamical systems (\ref{Eq:latent_lds}) featuring no, low, or high non-normality (Fig.~\ref{fig: fig1_setup}c).
We kept the recurrent self-input strengths constant, fixing the eigenvalues of $\mathbf{A}$ at $[0.7, 0.8, 0.9]$. By then varying the inter-node weights, we altered the angles between the eigenvectors to control the system's non-normality (see Appendix~\ref{app:nonnormal}).

For each AL method and system, the effects of the chosen stimulations on the system and subsequent data informativeness were evaluated.
At each time step $t$ for a given simulated neural network, a discrete linear system $\mathbf{y}(t+1) = \hat{\mathbf{W}} \mathbf{y}(t) + \hat{\mathbf{B}} \mathbf{u}(t)$ using least squares regression was fitted to the history of stimulations and neural responses (see Appendix~\ref{app:fitting}).
The fitted system was then used by the AL method to choose the next optimal input $\mathbf{u}(T)$, where $T$ indexes the stimuli (Appendix~\ref{app: stimulation_choice}).
This stimulation was applied to the simulated neural network, which returned the resulting response over the next $T_{trial}=300$ time steps $[\mathbf{y}(T), \mathbf{y}(T+1), \ldots, \mathbf{y}(T+T_{trial})]$ (Appendix~\ref{app:comp_experiments}).
The stimulation trial length was chosen to be considerably longer than the timescale of the dynamics $\tau$ in (\ref{Eq:latent_lds}), such that the system returns to resting state by the end of the stimulation trial.
This procedure was repeated for $K=1000$ trials, resulting in an estimated system after each perturbation.
We used the true projection matrix $\mathbf{J}$ to project $\hat{\mathbf{W}}$ back into the true latent space to estimate their respective $\hat{\mathbf{A}}=\mathbf{J}\hat{\mathbf{W}}\mathbf{J}^T$. This post-hoc evaluation separates the latent manifold geometry estimation from the latent dynamics estimation, focusing on the latter.

Each of the 16 possible combinations of the methodological assumptions in Fig.~\ref{fig: fig1_setup}a was evaluated (Appendix~\ref{app:system_eval}). To prevent the AL methods from getting stuck in a local minima, we used an $\epsilon$-greedy strategy: the AL strategy was used for 80\% of stimulations, while a randomly selected single neuron was stimulated in the remaining 20\%.
These methods were compared to a purely random stimulation baseline (`random'), and two oracle baselines (`Balanced GT EVD' and `Balanced GT EVD' in the Appendix plots). The oracle methods used the \emph{known} ground truth system to choose uniformly among either the top three right singular vectors or left eigenvectors of $\mathbf{W}$ for each stimulation (see Appendix~\ref{app: stimulation_choice}).

\section{Fastest active learning is achieved by minimizing the estimation error of a low-rank recurrent dynamics matrix}

\begin{figure}[t]
  \centering
  \includegraphics[width=0.8\linewidth]{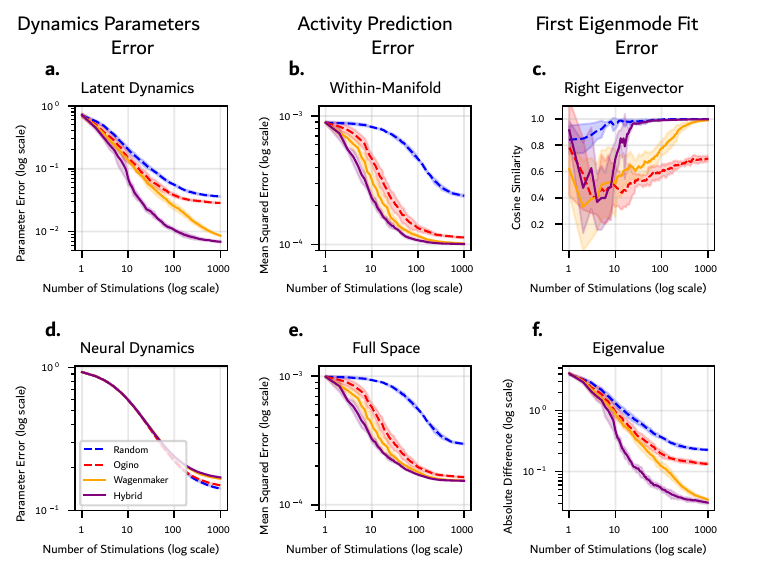}
  \caption{Low-rank dynamics models (solid) identify within-manifold dynamics faster than full-rank models (dashed).
    \textbf{a.} An estimation error for latent connectivity $\mathbf{A}$.
    \textbf{b.} Prediction error for latent dynamics.
    \textbf{c.} Alignment of the first eigenvectors to the GT.
    \textbf{d-e.} Similar to (\textbf{a-b}), but for the full connectivity matrix $\mathbf{W}$ and neural activity in the full space.
    \textbf{f.} Estimation error for the first eigenvalue of $\mathbf{A}$.}
  \label{fig: fig2_manifold}
\end{figure}

To determine which modelling choices in Fig.~\ref{fig: fig1_setup}a drive the most effective system identification, we evaluated the methods based on their ability to identify the recurrent dynamics parameter (parameter error: $\| \mathbf{W} - \hat{\mathbf{W}} \| / \| \mathbf{W} \|$) and predicted activity (Mean Squared Error), both for the latent and full spaces of a fully observable, normal system.
Fig.~\ref{fig: fig2_manifold} demonstrates a clear trade-off: prioritizing manifold identification sacrifices full-system accuracy, and vice versa.
By targeting fast-decaying terms, the objective of \citet{ogino_designing_2025} diverts stimulation resources to off-manifold directions.
This leads to poor latent dynamics estimation (dashed lines in Fig.~\ref{fig: fig2_manifold}a, red ~\citet{ogino_designing_2025} is similar to the blue `random' baseline), misaligned with the latent manifold (Fig.~\ref{fig: fig2_manifold}c), but yields the best performance for the full system estimation (Fig.~\ref{fig: fig2_manifold}d).
This excellent estimation of the full-system parameters $\mathbf{W}$ does not translate into precision of neural activity prediction (Fig.~\ref{fig: fig2_manifold}e), which is dominated by the on-manifold dynamics.
In contrast, the \citet{wagenmaker_active_nodate} objective sacrifices accuracy of the full system identification to prioritize learning latent dynamics. It fails to learn the geometry, i.e. the manifold embedding in the full space (Fig.~\ref{fig: fig2_manifold}c), yet better estimates the within-manifold dynamics compared to the full rank methods (Fig.~\ref{fig: fig2_manifold}f).

The method that yields the fastest and most accurate identification of the latent dynamics -- and therefore the activity prediction -- is a hybrid of the approaches in~\citet{wagenmaker_active_nodate} and~\citet{ogino_designing_2025}.
This method, named `Hybrid', sets $\mathbf{P}$ in (2) equal to the top 3 left eigenvectors (or singular vectors -- they are equivalent for a normal system) of $\mathbf{W}$. We used the following modeling choices (Fig.~\ref{fig: fig1_setup}a): the causal connectivity matrix is $\mathbf{W}$, the system rank is low, system modes are either EVD or SVD, and the inputs have isotropic covariance. By prioritizing isotropic stimulation on the manifold, this strategy achieves a significant increase in performance compared to both~\citet{wagenmaker_active_nodate} and~\citet{ogino_designing_2025}.
`Hybrid' is the most efficient in identifying the latent system across all of the 16 possible combinations as can be seen in Appendix~\ref{app:all_hybrid_methods}.
The advantage of estimating recurrent matrix $\mathbf{W}$ compared to the causality matrix $\mathbf{H}$ is likely due to its independence of the unknown actuation matrix $\mathbf{B}$.
Hence, if $\mathbf{B}$ is misidentified, even indirectly in a process of fitting $\mathbf{H}$ to the stimulation-activity response, this can lead to the directions of the strongest activity amplification, i.e. left eigenvectors, not being stimulated by the targeted input.
Indeed, as can be seen in Fig.~\ref{fig: fig2_manifold}c and Fig.~\ref{fig: fig2_manifold}f, using Wagenmaker's method leads to an efficient identification of the eigenvalues of $\mathbf{A}$ (up to the discrete-time approximation; see Appendix~\ref{app:system_eval}) but requires more stimulation steps to align with the true eigenvectors.

\section{Avoiding orthogonal inputs is essential to identify latent non-normal dynamics}

Evaluation on the normal system has shown that a hybrid of \citet{wagenmaker_active_nodate} and \citet{ogino_designing_2025} is the most effective for identifying on-manifold dynamics.
For a non-normal system, the geometry of their eigendecomposition presents additional modeling choices.
First, left and right eigenvectors do not align (Fig.~\ref{fig: fig1_setup}c.ii-iii), and the EVD and SVD (Fig.~\ref{fig: fig1_setup}a.iii) yield different bases for a non-normal system.
The left eigenvectors $\mathbf{q}_l$ are not orthogonal to each other and align with the input directions for the eigenmodes of the system $\mathbf{W}$, while the right singular vectors $\mathbf{v}_r$ define the orthonormal basis in the input space of $\mathbf{W}$.
Thus, we tested whether utilizing the top left eigenvectors $\mathbf{q}_l$, as opposed to right singular vectors $\mathbf{v}_r$, as the basis for stimulation inputs makes a difference for the low and high non-normal dynamics (Fig.~\ref{fig: fig1_setup}a.iii).
Note that since $\mathbf{q}_l$ are not orthogonal, the assumption on independence of the input modes does not hold.

\begin{figure}[h!]
  \centering
  \includegraphics[width=0.8\linewidth]{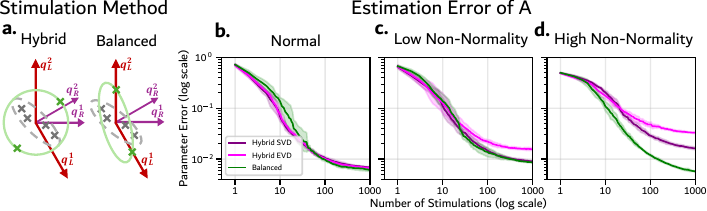}
  \caption{In non-normal systems, balanced stimulation of estimated left eigenvectors is more efficient in latent system identification than any Hybrid of AL objectives (\ref{Eq:optimisation_al_error}) derived in earlier works.
    \textbf{a.} While the Hybrid method stimulates directions of low input variance, the Balanced method sets the next stimulation (green crosses) along one of the left eigenvectors $\mathbf{q}_L$.
    Grey crosses show previous stimulations
    and the grey ellipse their resulting covariance.
    \textbf{b}-\textbf{d}.~Estimation error of the latent dynamics parameters $\mathbf{A}$ for (\textbf{b}) no, (\textbf{c}) low, and (\textbf{d}) high non-normality.}
  \label{fig: fig3_nonormality}
\end{figure}

To investigate whether minimizing (\ref{Eq:optimisation_al_error}) is still the most efficient AL approach for non-normal systems, we proposed one additional method: `Balanced'. The difference between Hybrid (using EVD) and Balanced is visualised in Fig.~\ref{fig: fig3_nonormality}a.
Given a history of stimulation inputs projected onto the left eigenvectors of a non-normal system (grey crosses) and their covariance matrix (grey circle), the Hybrid method aims to make the covariance matrix more isotropic (green circle) by stimulating along the direction of the smallest variance (green crosses) to minimise the system estimation error (\ref{Eq:optimisation_al_error}).
In contrast, at each stimulation point, the Balanced method uniformly chooses to stimulate one of the top three left eigenvectors $\mathbf{q}_l$ of the estimated recurrent weight matrix $\hat{\mathbf{W}}$. Since each of the eigenmodes are chosen with equal probability, the stimulation `budget' is `Balanced'.
These stimulation targets (green crosses in Fig.~\ref{fig: fig3_nonormality}a) lead to an anisotropic covariance matrix (green ellipse).

We tested our Hybrid method using either EVD or SVD and our Balanced method on identifying the latent dynamic system for different levels of non-normality (Fig.~\ref{fig: fig3_nonormality}b-d; see Appendix~\ref{app:all_hybrid_methods} for performances of the other model variants). For the normal case, both Hybrid methods show equivalent performance. Balanced starts learning with a slightly higher error, but approaches Hybrid for a large number of stimulations. This aligns with the observation that in a normal system, both objectives lead to spherical input covariance (green ellipsoid is a sphere in Fig.~\ref{fig: fig3_nonormality}a).
While their asymptotic behavior is the same,
Hybrid is more effective early in low-sample regime due to the systems eigenvectors not being identified yet.
In contrast, for the non-normal cases, Hybrid performs considerably worse than Balanced, failing to converge on the latent dynamics matrix $\mathbf{A}$, while the Balanced method requires only 19~stimulations to reach the error level otherwise achieved in 1000~random stimulations (Table~\ref{table1}). This difference in parameter estimation appears to stem from a misidentification of the eigenvalues and not the eigenvectors of the system (see Appendix~\ref{apd_eigvalsvsvec}).

This empirical result is formally supported by an analytical derivation (see Appendix~\ref{app:analytical}). By evaluating the true A-optimality objective, i.e. minimizing the trace of the inverse Fisher Information for a highly non-normal system (as non-normality $k \to \infty$), we have analytically compared the asymptotic error scalings of different methods.
We show that our Balanced method achieves an error scaling of $\frac{2}{2-\epsilon} \sum (-\lambda_i)$, which is strictly lower than the bounds achieved by the~\citet{wagenmaker_active_nodate} ($2 \sum (-\lambda_i)$) and~\citet{ogino_designing_2025} ($\frac{2}{\epsilon} \sum (-\lambda_i)$) objectives (given $0<\varepsilon<1$). This provides an analytical proof by counterexample that the approximation in (\ref{Eq:optimisation_al_error}) is no longer A-optimal for non-normal systems: providing correlated stimulation inputs along non-orthogonal left eigenvectors utilizes transient amplification to yield the tightest constraint on the parameter estimation error.

\begin{table}
  \small
  \centering
  \setlength{\tabcolsep}{3pt}
  \begin{threeparttable}
    \caption{Model assumptions and performance measured as the number of stimulations (mean $\pm$ SD, 5 seeds) required to reach an error threshold of the weakest baseline model: `random' for learning $\mathbf{A}$, and `GT' -- for learning $\mathbf{W}$ (\textbf{lower} is better).}
    \label{table1}

    \begin{tabular}{@{}l cccc cccccc@{}}
      \toprule
      \multicolumn{5}{c}{~} & \multicolumn{6}{c}{Performance (\#stim)}                                                                                                                                                                                                                                        \\
      \cmidrule(r){6-11}
      ~                     & \multicolumn{4}{c}{Model assumptions}    & \multicolumn{2}{c}{Normal} & \multicolumn{2}{c}{Low non\nobreakdash-norm.} & \multicolumn{2}{c}{High non\nobreakdash-norm.}                                                                                                          \\
      \cmidrule(r){2-5} \cmidrule(r){6-7} \cmidrule(r){8-9} \cmidrule(r){10-11}
      Name                  & $\mathbf{M}$                             & rank                       & mode                                          & isotrop                                        & $\mathbf{A}$      & $\mathbf{W}$ & $\mathbf{A}$      & $\mathbf{W}$ & $\mathbf{A}$      & $\mathbf{W}$ \\
      \midrule
      Wagenmaker\tnote{a}   & $\mathbf{H}$                             & low                        & SVD                                           & input                                          & $56\pm13$         & $86\pm2$     & $93\pm23$         & $87\pm2$     & $116\pm57$        & $83\pm1$     \\
      Ogino\tnote{b}        & $\mathbf{W}$                             & full                       & EVD                                           & activity                                       & $116\pm22$        & $74\pm1$     & $199\pm75$        & $73\pm1$     & $313\pm114$       & $74\pm1$     \\
      \textbf{Hybrid}       & $\mathbf{W}$                             & low                        & SVD                                           & input                                          & $\mathbf{16\pm3}$ & $87\pm1$     & $\mathbf{24\pm7}$ & $87\pm1$     & $42\pm6$          & $88\pm1$     \\
      \textbf{Balanced}     & $\mathbf{W}$                             & low                        & EVD                                           & approx.                                        & $25\pm4$          & $189\pm13$   & $25\pm3$          & $186\pm12$   & $\mathbf{19\pm4}$ & $193\pm12$   \\
      \bottomrule
    \end{tabular}
    \begin{tablenotes}
      \footnotesize
      \item[] \makebox[0.5\linewidth][l]{$^a$ \citet{wagenmaker_active_nodate}}%
      \makebox[0.5\linewidth][l]{$^b$ \citet{ogino_designing_2025}}
    \end{tablenotes}

  \end{threeparttable}
\end{table}

\section{Conclusion and Discussion}

This paper presents a unified mathematical framework and benchmark to evaluate AL for identifying neural population dynamics under challenging, biologically relevant assumptions. We revealed how underlying modelling choices impact the speed and accuracy of system identification by systematizing state-of-the-art methods into four key dimensions:
model parametrization (state transition or stimulus-response), system rank (low or full), dynamics modes (singular or eigen), and input sampling strategies (isotropic input or response).

Our evaluation yielded two critical insights for experimental design.
First, we demonstrated a clear trade-off between full and latent system identification. Prioritizing the identification of the low-dimensional manifold inherently sacrifices accuracy in identifying the full system, and vice versa (Fig.~\ref{fig: fig2_manifold}). Full-rank models utilize increasingly large~\citep{ogino_designing_2025} stimulation resources on off-manifold directions to capture fast-decaying terms, whereas low-rank models~\citep{wagenmaker_active_nodate} focus on the latent dynamics that are the central source of neural response variability. Given this, we proposed a new hybrid approach which prioritizes the identification of the top modes of the recurrent matrix, yielding more accurate neural activity predictions than either of the two models.
Second, methods must explicitly account for the biologically relevant features of neuronal network connectivity, particularly their non-normal dynamics induced by functional circuits or Dale's law~\citep{kumar2022non}. Traditional approaches relying on orthogonality of dynamic eigenmodes fail to capture transient amplifications. We demonstrated with a novel `Balanced' method that leveraging low-rank dynamics through eigendecomposition to estimate the recurrent matrix can more efficiently identify latent non-normal systems (Fig.~\ref{fig: fig3_nonormality}).

Despite these advances, several limitations in our current framework must be acknowledged.
Our benchmark assumes 1) linear dynamical systems 2) optimized via A-optimality.
While limited to linear systems, our analysis serves as a foundational step for analyzing non-linear neural dynamics (see Appendix~\ref{app:non-linear}). In the future work, these AL strategies can be extended to Conditionally Linear Dynamical Systems~\citep{geadah2026modeling} or applied via local linearizations along a neural state trajectory.
Similarly, our results assume full observability and controllability, which is not the case in experimental settings (see Appendix~\ref{sec:partialobservation} for the impact of partial observability on system identification).Deriving tighter approximations for the error term under non-normality remains an important direction for future theoretical work. Finally, alternative optimization criteria (such as T-optimality), non-Gaussian observation models, or even implicit models with likelihood-free inference techniques~\citep{kleinegesse2020bayesian}, biological safety, and computational latency of real-time model refitting remain important future directions.

The strength of our benchmark lies in revealing how different modeling choices recover ground truth dynamics; however, evaluation on experimental data is a critical next step. The scarcity of open perturbation datasets across diverse brain areas, recording modalities, and species remains a significant obstacle to constructing rigorous resampling-based pseudo-experiments~\citep{wagenmaker_active_nodate}.
Ultimately, optimal experimental design is highly context-dependent; methodological choices must be tailored to the system's actual latent structure and the study's specific goals. As closed-loop neuroscience advances, addressing these questions will be vital for unlocking the full potential of targeted neural manipulation.

\acks{We thank Angus Chadwick for insightful discussions and for highlighting the work of \cite{ogino_designing_2025}.

  This project was supported by the United Kingdom Research and Innovation (grant EP/Y030869/1), UKRI AI Centre for Doctoral Training in Biomedical Innovation at the University of Edinburgh.
  NK was supported by the Royal Society University Research Fellowship (URF/R1/241060).
  For the purpose of open access, the author has applied a Creative Commons Attribution (CC BY) licence to any Author Accepted Manuscript version arising from this submission.}

\bibliography{references}

\begin{thebibliography}{20}
\providecommand{\natexlab}[1]{#1}
\providecommand{\url}[1]{\texttt{#1}}
\expandafter\ifx\csname urlstyle\endcsname\relax
  \providecommand{\doi}[1]{doi: #1}\else
  \providecommand{\doi}{doi: \begingroup \urlstyle{rm}\Url}\fi

\bibitem[Bolus et~al.(2021)Bolus, Willats, Rozell, and
  Stanley]{bolus_state-space_2021}
Michael~F Bolus, Adam~A Willats, Christopher~J Rozell, and Garrett~B Stanley.
\newblock State-space optimal feedback control of optogenetically driven neural
  activity.
\newblock \emph{Journal of Neural Engineering}, 18\penalty0 (3):\penalty0
  036006, 2021.

\bibitem[Bondanelli and
  Ostojic(2020)]{bondanelliCodingTransientTrajectories2020}
Giulio Bondanelli and Srdjan Ostojic.
\newblock Coding with transient trajectories in recurrent neural networks.
\newblock \emph{PLOS Computational Biology}, 16\penalty0 (2):\penalty0
  e1007655, 2020.

\bibitem[Gallego et~al.(2018)Gallego, Perich, Naufel, Ethier, Solla, and
  Miller]{gallego2018cortical}
Juan~A Gallego, Matthew~G Perich, Stephanie~N Naufel, Christian Ethier, Sara~A
  Solla, and Lee~E Miller.
\newblock Cortical population activity within a preserved neural manifold
  underlies multiple motor behaviors.
\newblock \emph{Nature Communications}, 9\penalty0 (1):\penalty0 4233, 2018.

\bibitem[Geadah et~al.(2025)Geadah, Nejatbakhsh, Lipshutz, Pillow, and
  Williams]{geadah2026modeling}
Victor Geadah, Amin Nejatbakhsh, David Lipshutz, Jonathan Pillow, and Alex
  Williams.
\newblock Modeling neural activity with conditionally linear dynamical systems.
\newblock \emph{Advances in Neural Information Processing Systems},
  38:\penalty0 87635--87665, 2025.

\bibitem[Goldman(2009)]{goldman2009memory}
Mark~S Goldman.
\newblock Memory without feedback in a neural network.
\newblock \emph{Neuron}, 61\penalty0 (4):\penalty0 621--634, 2009.

\bibitem[Hennequin et~al.(2012)Hennequin, Vogels, and
  Gerstner]{hennequin2012non}
Guillaume Hennequin, Tim~P Vogels, and Wulfram Gerstner.
\newblock Non-normal amplification in random balanced neuronal networks.
\newblock \emph{Physical Review E—Statistical, Nonlinear, and Soft Matter
  Physics}, 86\penalty0 (1):\penalty0 011909, 2012.

\bibitem[Kleinegesse and Gutmann(2020)]{kleinegesse2020bayesian}
Steven Kleinegesse and Michael~U Gutmann.
\newblock Bayesian experimental design for implicit models by mutual
  information neural estimation.
\newblock In \emph{International Conference on Machine Learning}, pages
  5316--5326. PMLR, 2020.

\bibitem[Kumar and Bouchard(2022)]{kumar2022non}
Ankit Kumar and Kristofer Bouchard.
\newblock Non-normality in neural networks.
\newblock In \emph{AI and Optical Data Sciences III}, volume 12019, pages
  70--76. SPIE, 2022.

\bibitem[Marshel et~al.(2019)Marshel, Kim, Machado, Quirin, Benson, Kadmon,
  Raja, Chibukhchyan, Ramakrishnan, Inoue, et~al.]{marshel2019cortical}
James~H Marshel, Yoon~Seok Kim, Timothy~A Machado, Sean Quirin, Brandon Benson,
  Jonathan Kadmon, Cephra Raja, Adelaida Chibukhchyan, Charu Ramakrishnan,
  Masatoshi Inoue, et~al.
\newblock Cortical layer--specific critical dynamics triggering perception.
\newblock \emph{Science}, 365\penalty0 (6453):\penalty0 eaaw5202, 2019.

\bibitem[Minai et~al.(2024)Minai, Soldado-Magraner, Smith, and
  Yu]{minai_miso_nodate}
Yuki Minai, Joana Soldado-Magraner, Matthew~A Smith, and Byron~M Yu.
\newblock {MiSO}: {Optimizing} brain stimulation to create neural population
  activity states.
\newblock \emph{Advances in Neural Information Processing Systems},
  37:\penalty0 24126--24149, 2024.

\bibitem[Murphy and Miller(2009)]{murphy2009balanced}
Brendan~K Murphy and Kenneth~D Miller.
\newblock Balanced amplification: a new mechanism of selective amplification of
  neural activity patterns.
\newblock \emph{Neuron}, 61\penalty0 (4):\penalty0 635--648, 2009.

\bibitem[Nejatbakhsh et~al.(2023)Nejatbakhsh, Fumarola, Esteki, Toyoizumi,
  Kiani, and Mazzucato]{nejatbakhsh2023predicting}
Amin Nejatbakhsh, Francesco Fumarola, Saleh Esteki, Taro Toyoizumi, Roozbeh
  Kiani, and Luca Mazzucato.
\newblock Predicting the effect of micro-stimulation on macaque prefrontal
  activity based on spontaneous circuit dynamics.
\newblock \emph{Physical Review Research}, 5\penalty0 (4):\penalty0 043211,
  2023.

\bibitem[Ogino et~al.(2026)Ogino, Sekizawa, Kitazono, and
  Oizumi]{ogino_designing_2025}
Mikito Ogino, Daiki Sekizawa, Jun Kitazono, and Masafumi Oizumi.
\newblock Designing optimal perturbation inputs for system identification in
  neuroscience.
\newblock \emph{eLife}, 15:\penalty0 RP110030, 2026.

\bibitem[O’Shea et~al.(2022)O’Shea, Duncker, Goo, Sun, Vyas, Trautmann,
  Diester, Ramakrishnan, Deisseroth, Sahani, et~al.]{o2022direct}
Daniel~J O’Shea, Lea Duncker, Werapong Goo, Xulu Sun, Saurabh Vyas, Eric~M
  Trautmann, Ilka Diester, Charu Ramakrishnan, Karl Deisseroth, Maneesh Sahani,
  et~al.
\newblock Direct neural perturbations reveal a dynamical mechanism for robust
  computation.
\newblock \emph{bioRxiv}, pages 2022--12, 2022.

\bibitem[Perich et~al.(2025)Perich, Narain, and Gallego]{perich2025neural}
Matthew~G Perich, Devika Narain, and Juan~A Gallego.
\newblock A neural manifold view of the brain.
\newblock \emph{Nature Neuroscience}, 28\penalty0 (8):\penalty0 1582--1597,
  2025.

\bibitem[Pukelsheim(2006)]{pukelsheim2006optimal}
Friedrich Pukelsheim.
\newblock \emph{Optimal design of experiments}.
\newblock SIAM, 2006.

\bibitem[Qian et~al.(2024)Qian, Zavatone-Veth, Ruben, and
  Pehlevan]{qian2024partial}
William Qian, Jacob~A Zavatone-Veth, Benjamin~S Ruben, and Cengiz Pehlevan.
\newblock Partial observation can induce mechanistic mismatches in
  data-constrained models of neural dynamics.
\newblock \emph{Advances in Neural Information Processing Systems},
  37:\penalty0 67467--67510, 2024.

\bibitem[Tentori et~al.(2025)Tentori, Kastellakis, Maschietto, Leparulo,
  Poirazi, Mazzucato, Allegra, and Vassanelli]{tentori2025spontaneous}
Elisa Tentori, George Kastellakis, Marta Maschietto, Alessandro Leparulo,
  Panayiota Poirazi, Luca Mazzucato, Michele Allegra, and Stefano Vassanelli.
\newblock Spontaneous dynamics predict the effects of targeted intervention in
  hippocampal neuronal cultures.
\newblock \emph{bioRxiv}, 2025.

\bibitem[Vyas et~al.(2020)Vyas, Golub, Sussillo, and
  Shenoy]{vyas2020computation}
Saurabh Vyas, Matthew~D Golub, David Sussillo, and Krishna~V Shenoy.
\newblock Computation through neural population dynamics.
\newblock \emph{Annual Review of Neuroscience}, 43\penalty0 (1):\penalty0
  249--275, 2020.

\bibitem[Wagenmaker et~al.(2024)Wagenmaker, Mi, Rozsa, Bull, Svoboda, Daie,
  Golub, and Jamieson]{wagenmaker_active_nodate}
Andrew Wagenmaker, Lu~Mi, Marton Rozsa, Matthew~S Bull, Karel Svoboda, Kayvon
  Daie, Matthew~D Golub, and Kevin Jamieson.
\newblock Active learning of neural population dynamics using two-photon
  holographic optogenetics.
\newblock \emph{Advances in Neural Information Processing Systems},
  37:\penalty0 31659--31687, 2024.

\end{thebibliography}

%%%%%%%%%%%%%%%%%%%%%%%%%%%%%%%%%%%%%%%%%%%%%%%%%%%%%%%%%%%%
%Appendix
%%%%%%%%%%%%%%%%%%%%%%%%%%%%%%%%%%%%%%%%%%%%%%%%%%%%%%%%%%%%

\clearpage
\appendix

\section{Analytical comparison of asymptotic optimality for an extremely non-normal system}
\label{app:analytical}
Before going into the derivation, we must emphasize that (\ref{Eq:optimisation_al_error}) looks deceptively similar to the A-optimality definition (\ref{Eq:optimisation_A_opt}), yet it is not identical:

A-optimality (w.r.t. $\mathbf{W}$):
\quad$\mathbf{u} = \arg\min \text{Tr}((\mathcal{I}(\mathbf{W}))^{-1})$

Eq 2:
~\qquad~\qquad~\qquad\qquad~$\mathbf{u} = \arg\min \text{Tr}\left( (\mathbf{P} \boldsymbol{\Sigma}_U \mathbf{P}^\top)^{-1} \right)$

Here $\mathcal{I}(W) = \sigma^{-2} \mathbf{\Sigma}_X$, where $\sigma$ is the magnitude of the Gaussian state transition noise $\mathcal{N}(0,\sigma I)$, and $\mathbf{\Sigma}_X$ is the empirical covariance of evoked neural responses.
Note, that~\citet{wagenmaker_active_nodate} is also NOT A-optimal w.r.t. $\mathbf{H}$, since in that case $\mathbf{\Sigma}_X = (\mathbf{H} u)(\mathbf{H} u)^\top = \mathbf{H}\Sigma_u \mathbf{H}^\top \neq \mathbf{V_r}\Sigma_u \mathbf{V_r}^\top$ (in general).

With the following derivation, we aim to prove that our heuristic `Balanced' method is closer to the A-Optimality objective than any currently proposed approximations. The idea of the proof is to show on a counterexample that these approximations in (2) yield sub-optimal solutions in terms of A-optimality, with our approach of stimulating along $\mathbf{Q}_l$ yielding lower $\text{Tr} (\mathcal{I}(\mathbf{W})^{-1})$ and thus lower error in parameter estimation.

The proof thus goes as follows. Consider a simple matrix
$$\mathbf{W} = \begin{bmatrix} \lambda_1 & k \\ 0 & \lambda_2\end{bmatrix}$$
governing stable dynamics with $k$ regulating non-normality. The tranference matrix is thus:
$$\mathbf{H} = \int_{0}^{T} e^{\mathbf{W}t} \, dt = \begin{bmatrix} \frac{e^{\lambda_1 T} - 1}{\lambda_1} & \frac{k}{\Delta} \left( \frac{e^{\lambda_1 T} - 1}{\lambda_1} - \frac{e^{\lambda_2 T} - 1}{\lambda_2} \right) \\ 0 & \frac{e^{\lambda_2 T} - 1}{\lambda_2} \end{bmatrix}$$
where $\Delta = \lambda_1 - \lambda_2$, and $\lambda_{1,2}<0$.

Then eigenvectors for $W$ are:
$$\mathbf{Q}_r = \begin{bmatrix} 1 & -\cfrac{k}{\sqrt{k^2+\Delta^2}} \\ 0 & \cfrac{\Delta}{\sqrt{k^2+\Delta^2}} \end{bmatrix}, \quad \mathbf{Q}_l = \begin{bmatrix} \cfrac{\Delta}{\sqrt{k^2+\Delta^2}} & 0 \\ \cfrac{k}{\sqrt{k^2+\Delta^2}} & 1 \end{bmatrix}$$
\emph{Note, vectors are in columns for both, so $\mathbf{Q}_l^T\mathbf{W} = \Lambda \mathbf{Q}_l^T$.}

And singular vectors for $\mathbf{H}$ have a complex closed-form solution, yet, crucially, it yields an orthonormal basis, and for both $k=0$ and $k\gg1$:
$$\mathbf{V}_l \approx \begin{bmatrix} 1 & 0 \\ 0 & 1 \end{bmatrix}$$

\subsection{Input covariance \texorpdfstring{$\mathbf{\Sigma}_u$}{Sigma u}}
Then, we need to calculate the input covariance $\mathbf{\Sigma}_u$. For our method it is simply:
$\mathbf{\Sigma}_{U,\textrm{ours}} = \mathbf{Q}_l\mathbf{Q}_l^\top = \frac{1}{k^2+\Delta^2} \begin{bmatrix} \Delta^2 & k\Delta \\ k\Delta & 2k^2+\Delta^2 \end{bmatrix}$
For Wagenmaker and Ogino we need to solve eq. (2):
$$ \mathbf{\Sigma}_{U,\text{Wag}} = \begin{bmatrix} 1 & 0 \\ 0 & 1 \end{bmatrix} \quad \mathbf{\Sigma}_{U,\text{Ogino}} \propto \left(\mathbf{Q}_r \mathbf{Q}_r^\top\right)^{-1/2}$$
We also analytically compare this to a method that theoretically achieves the maximal response energy via transient amplification:
$$\mathbf{\Sigma}_{U_,\textrm{obsG}} = \mathbf{G}_{\text{obs}}$$
where $\mathbf{G}_{\text{obs}}$ is the infinite-time observability Gramian, computed as the solution to the Lyapunov equation:
$$\mathbf{W}^\top \mathbf{G}_{\text{obs}} + \mathbf{G}_{\text{obs}} \mathbf{W} = -\mathbf{I}$$
We normalize all methods have the same stimulation `budget' $\text{Tr} (\mathbf{\Sigma}_u) = 2$.

Finally, we add the same exploration term for each method:
$$\mathbf{\Sigma}_U = \epsilon \mathbf{I} + (1-\epsilon)\mathbf{\Sigma}_{U,{\textrm{method}}}$$
where $\epsilon=0.2$.

\subsection{Response covariance \texorpdfstring{$\mathbf{\Sigma}_X$}{Sigma X}}
Once $\mathbf{\Sigma}_U$ is determined, the true evoked response covariance $\mathbf{\Sigma}_X$ must be evaluated over the entire continuous trajectory:
$$\mathbf{\Sigma}_X = \int_0^\infty \mathbf{x}(t)\mathbf{x}(t)^\top dt$$
This integral is the exact analytical solution to the continuous Lyapunov equation: $\mathbf{W} \mathbf{\Sigma}_X + \mathbf{\Sigma}_X \mathbf{W}^\top = -\mathbf{\Sigma}_U$. We then compute the true A-optimality objective $\propto \text{Tr}(\mathbf{\Sigma}_X^{-1})$.

Finally, we summarize the Fisher information for each of the approaches for a system with $\lambda_1=-2$ and $\lambda_2=-1$:

\begin{table}[h]
  \centering
  \caption{A-Optimality for highly non-normal 2D system across methods}
  \label{tab:normalized_fisher_info}
  \renewcommand{\arraystretch}{2.5}
  \resizebox{\textwidth}{!}{
    \begin{tabular}{lccccc}
      \toprule
      \textbf{Strategy}                                                                                                                                     & \textbf{Input Cov. ($\boldsymbol{\Sigma}_U$)}                                                            & \textbf{Fisher Info. $\propto \text{Tr}(\mathbf{\Sigma}_X)$} & \textbf{Evoked Det. $\det(\mathbf{\Sigma}_X)$}      & \textbf{A-Optimality $\propto \text{Tr}(\mathbf{\Sigma}_X^{-1})$} & \textbf{Asymptote ($k \to \infty$)}                  \\
      \midrule
      Wagenmaker                                                                                                                                            &
      $\mathbf{I}$                                                                                                                                          &
      $\cfrac{k^2+9}{12}$                                                                                                                                   &
      $\cfrac{k^2+9}{72}$                                                                                                                                   &
      $6$                                                                                                                                                   &
      $-2(\lambda_1+\lambda_2) = 6$                                                                                                                                                                                                                                                                                                                                                                                                                                                                                    \\
      Ogino ($\mathbf{P}=\mathbf{W}$)                                                                                                                       &
      $\epsilon\mathbf{I} + \cfrac{2(1-\epsilon)\left(\mathbf{W}^\top \mathbf{W}\right)^{-1/2}}{\text{Tr}\left((\mathbf{W}^\top \mathbf{W})^{-1/2}\right)}$ &
      $\cfrac{k^4 + 122k^2 + 441}{60(k^2+9)}$                                                                                                               &
      $\cfrac{k^6 + 163k^4 + 4291k^2 + 16929}{1800(k^2+9)^2}$                                                                                               &
      $\cfrac{30(k^2+9)(k^4 + 122k^2 + 441)}{k^6 + 163k^4 + 4291k^2 + 16929}$                                                                               &
      $\cfrac{-2(\lambda_1+\lambda_2)}{\epsilon} = 30$                                                                                                                                                                                                                                                                                                                                                                                                                                                                 \\
      Max amplification ($\mathbf{G}_{\text{obs}}$)                                                                                                         & $\epsilon \mathbf{I} + \frac{2(1-\epsilon)}{\text{Tr}(\mathbf{G}_{\text{obs}})} \mathbf{G}_{\text{obs}}$ & $\cfrac{9k^2+49}{60}$                                        & $\cfrac{81k^6+963k^4+5011k^2+16929}{1800(k^2+9)^2}$ & $\cfrac{30(9k^2+49)(k^2+9)^2}{81k^6+963k^4+5011k^2+16929}$        & $\cfrac{-2(\lambda_1+\lambda_2)}{2-\epsilon} = 10/3$ \\
      Our `Balanced' ($\mathbf{Q}_l$)                                                                                                                       &
      $\epsilon\mathbf{I} + (1-\epsilon)\mathbf{Q}_l \mathbf{Q}_l^\top$                                                                                     &
      $\cfrac{9k^2+45}{60}$                                                                                                                                 &
      $\cfrac{9k^6+27k^4+43k^2+25}{200(k^2+1)^2}$                                                                                                           &
      $\cfrac{30(k^2+5)(k^2+1)^2}{9k^6+27k^4+43k^2+25}$                                                                                                     &
      $\cfrac{-2(\lambda_1+\lambda_2)}{2-\epsilon} = 10/3$                                                                                                                                                                                                                                                                                                                                                                                                                                                             \\
      \bottomrule
    \end{tabular}
  }
  \vspace{1ex}
  \raggedright
  {\small \textit{Note:} For the finite $k$ calculations, expressions are evaluated at exploration parameter $\epsilon=0.2$ and eigenvalues $\lambda_1=-2$ and $\lambda_2=-1$.}
\end{table}

Using SageMath for simplifying polynomials, we demonstrate that asymptotic A-optimality ($k\to\infty$) is $2\sum(-\lambda_i)$ for Wagenmaker, $\frac2{\varepsilon}\sum(-\lambda_i)$ for Ogino.
In contrast, both our `Balanced' method and the transient amplification baseline achieve a limit of $\frac2{(2-\varepsilon)}\sum(-\lambda_i)$.
For stable dynamics ($\sum(-\lambda_i)>0$), our method therefore yields a lower A-optimality target.
Since lower scores are better, this confirms that our approach is closer to the theoretical optimum.
The result holds for a 3-node feedforward chain, with the gap between methods widening (always $2\sum(-\lambda_i)$ for Wagenmaker, yet $\frac2{(3-2\varepsilon)}\sum(-\lambda_i)$ for our method).

We thus theoretically verify the physics intuition we provided in the manuscript, that stimulating left eigenvectors utilizes transient input amplification, yielding stronger responses ($\text{Tr}(\mathbf{\Sigma}_X)$) and a fast heuristic closer to an A-optimal solution w.r.t. $\mathbf{W}$, better constraining the network connectivity parameters.

\section{System set-up and fitting}
\subsection{Non-normal system}
\label{app:nonnormal}
A dynamical system is defined as non-normal when its recurrent matrix $\mathbf{W}$: $\mathbf{W} \mathbf{W}^\top \neq \mathbf{W}^\top\mathbf{W}$. Here, we measured the non-normality of a system as $\frac{\|\mathbf{W} \mathbf{W}^\top - \mathbf{W}^\top\mathbf{W}\|}{\|\mathbf{W} \mathbf{W}^\top\|}$, which is 0 if all eigenvectors of the system are orthogonal and upper bounded by $\sqrt{2}$ if all eigenvectors were perfectly aligned. Hence by setting the eigenvalues and angles between the right eigenvectors one can determine the non-normality level of the system. This was accomplished for the latent matrix $\mathbf{A} \in \mathbb{R}^{R \times R}$ of rank  $R=3$ by defining it as:

\[
  \mathbf{A} = \begin{bmatrix}
    \lambda_1 & w_1       & 0         \\
    0         & \lambda_2 & w_2       \\
    0         & 0         & \lambda_3
  \end{bmatrix},
\]

where $[\lambda_1, \lambda_2,\lambda_3]$ are the desired eigenvalues. For a normal system, the off-diagonal terms $w_1$ and $w_2$  can then simply be set to 0, so that the matrix is symmetric. For any other non-normal system, the desired angles $\alpha_1$ between the first and second right eigenvector and $\alpha_2$ between the first and third right eigenvector can be used to choose the feedforward weights:

\begin{equation}
  w_1 = \frac{\cos^2(\alpha_1)(\lambda_2-\lambda_1)}{\sqrt{1-\cos^2(\alpha_1)}} \quad \text{and} \quad
  w_2 = \frac{\cos^2(\alpha_2)(\lambda_3-\lambda_2)(\lambda_3-\lambda_1)}{\sqrt{w_1^2-w_1^2\cos^2(\alpha_2)-(\lambda_3-\lambda_1)^2 \cos^2(\alpha_2)}},
\end{equation}

which are positive and solvable as long as both
\[
  0^\circ < \alpha_1 < 90^\circ \quad \text{and} \quad
  \arccos(+\frac{w_1}{\sqrt{w_1^2+(\lambda_3-\lambda_1)^2}}) < \alpha_2 < \arccos(-\frac{w_1}{\sqrt{w_1^2+(\lambda_3-\lambda_1)^2}}).
\]

Here $\alpha_1$ and $\alpha_2$ were either $(90^\circ,90^\circ)$ for a normal system, $(40^\circ,60^\circ)$ for a system with low non-normality of 0.14, and $(1^\circ,2^\circ)$ for a system with high non-normality of 1.04. Note that for the non-normal case, the left and right eigenvectors corresponding to the same eigenvalue are not identical anymore, but sit at an angle to each other, with the left eigenvector being the inverse of the right.

To embed this latent feedforward network in the higher dimensional neural space of size $N=100$, the projection matrix $\mathbf{J} \in \mathbb{R}^{R \times N}$ was chosen to be semi-orthogonal, hence preserving the angles and eigenvalues for the top $R=3$ eigenvectors of $\mathbf{W}=\mathbf{J}^\top\mathbf{A}\mathbf{J} \in \mathbb{R}^{N \times N}$ with the other eigenmodes being orthogonal to the dynamics manifold, and therefore falling into the null-space of the transition matrix $\mathbf{W}$, formally corresponding to eigenvalue of 0. Since the full continuous neural system is defined as $\tau \frac{d\mathbf{x}}{dt} = \mathbf{W} \mathbf{x}(t) - \mathbf{x}(t) + \mathbf{B}\mathbf{u}(t) + \mathbf{w}(t) $ with $\tau=5$, this results in three slow eigenmodes and 97 fast ones.

\subsection{Computational experiments}
\label{app:comp_experiments}
The discretised version of the system was then used to generate the readout activity
by propagating $\mathbf{x}(t)$ and then observing
$\mathbf{y}(t)$ at each timepoint:
$$\mathbf{x}(t)= \mathbf{W}_d \mathbf{x}(t-1) + \mathbf{B}_d \mathbf{u}(t-1) + \mathbf{w}_d (t)   \in \mathbb{R}^{N} \text{ and }
  \mathbf{y}(t) = \mathbf{C}_d \mathbf{x}(t)  + \mathbf{v}_d(t) \in \mathbb{R}^{I} ,$$
where $\mathbf{W}_d=e^{\frac{dt}{\tau}(\mathbf{W}-\mathbf{I})}$ with
a time step of $dt=1$ and $\sigma^2=0.0001$ for both the discrete
dynamics noise
$\mathbf{w}_d\sim \mathcal{N}(\mathbf{0}, \sigma^2 \mathbf{I})$ and
readout noise
$\mathbf{v}_d\sim \mathcal{N}(\mathbf{0}, \sigma^2 \mathbf{I})$. The
readout matrix $\mathbf{C}_d \in \mathbb{R}^{I \times N}$ is set to
have orthonormal rows and to be the transpose of
$\mathbf{B}_d \in \mathbb{R}^{N \times I}$, i.e.
$\mathbf{B}_d = \mathbf{C}_d^\top$. The latent system (\ref{Eq:latent_lds}) could be discretised in the same way with  $\mathbf{A}_d=e^{\frac{dt}{\tau}(\mathbf{A}-\mathbf{I})}$.

$K=1000$ stimulations trials were carried out, with each stimulation trial being $T_{trial}=300$ time steps long, leading to a total number of $T = K*T_{trial} = 300 \ 000$ samples per simulation run. For each stimulation trial, input $\mathbf{u}(t=k*T_{trial}) \in \mathbb{R}^{I}$ ($k \in [0,\dots,K-1]$) was chosen by the AL method, and then set to 0 for the remainder of the time points $t \in [0, \dots, T]$, allowing neural activity to decay back to resting state before the next stimulation began.

All computational experiments were executed on an Intel(R) Xeon(R) W-2223 CPU (3.60 GHz). A full simulation run with $K=1000$ stimulations and $N=100$ neurons for one specific AL method took between 600 and 900 seconds (depending on the method, e.g., Balanced is faster than Hybrid), utilizing approximately 10~Gb of RAM.
Therefore, the final results plots with 16 hybrid AL methods, the Balanced method, two GT and one random baseline across 5 seeds each took approximately $20 \times 5 \times 800 = 80000 \mathrm{s} = 22 \mathrm{h}$.

\subsection{LDS fitting}
\label{app:fitting}
The input at each stimulation trial $k$ was chosen by the AL method. This was accomplished by first fitting a discrete linear dynamical system to the data gathered up to timepoint $t=kT_{trial} \in [T_{trial},2T_{trial} \dots, KT_{trial}]$ using Ridge Regression:

\begin{equation}
  [\hat{\mathbf{W}},\hat{\mathbf{B}}] = \mathbf{Y}_{t+1} \mathbf{Z}^\top \left(  \mathbf{Z}  \mathbf{Z}^\top + \lambda_W  \mathbf{I} \right)^{-1},
\end{equation}

with

\[
  \mathbf{Z} = \begin{bmatrix}
    \mathbf{y}(0) & \mathbf{y}(1) & \dots & \mathbf{y}(t-1) \\
    \mathbf{u}(0) & \mathbf{u}(1) & \dots & \mathbf{u}(t-1)
  \end{bmatrix} \in \mathbb{R}^{(N+I) \times t-1}, \quad \text{and}
\]
\[
  \mathbf{Y}_{t+1} =  \begin{bmatrix}\mathbf{y}(1) & \mathbf{y}(2) & \dots & \mathbf{y}(t) \end{bmatrix} \in \mathbb{R}^{N \times t-1}.
\]

Additionally, the transference matrix $\mathbf{H}$ was estimated using reduced rank regression by first fitting its least square solution for the inputs $\mathbf{U} = [\mathbf{u}(0),\mathbf{u}(T_{trial}),\dots,\mathbf{u}((k-1)T_{trial})]$, and the total recorded activity across each stimulation trial

$\mathbf{Y}_{\text{sum}} = [\sum_{t=0}^{T_{trial}} \mathbf{y}(t),\sum_{t=T_{trial}}^{2T_{trial}} \mathbf{y}(t),\dots,\sum_{t=(k-1)T_{trial}}^{kT_{trial}} \mathbf{y}(t)]$:

\[
  \hat{\mathbf{H}}_{LS}= \mathbf{Y}_{\text{sum}} \mathbf{U}^\top \left(  \mathbf{U}  \mathbf{U}^\top + \lambda_H  \mathbf{I} \right)^{-1} \in \mathbb{R}^{N \times I}
\]

and then reducing it to rank 3 by using its top 3 left singular vectors $\mathbf{v}$:

\[
  \hat{\mathbf{H}} = \sum_{i=1}^{r} \mathbf{v}_i \mathbf{v}_i^\top \hat{\mathbf{H}}_{LS}.
\]

The regression parameters were set to $\lambda_W=1$ and $\lambda_H=5$ for fitting $\hat{\mathbf{W}}$ and $\hat{\mathbf{H}}$ respectively.

\subsection{Stimulation choice}
\label{app: stimulation_choice}

For the 16 different combinations (all combinations in Fig.~\ref{fig: fig1_setup}), the fitted system was then used to set the causality matrix $\mathbf{M}$ as either $\hat{\mathbf{W}}$ or $\hat{\mathbf{H}}$, leaving it either as is or reducing it to its top $d=3$ modes based on its SVD or EVD where the modes are ordered from largest to smallest singular value or eigenvalue respectively. Either the full causality matrix $\mathbf{M}$ or just its right singular vectors / left eigenvectors were then used for the projection matrix $\mathbf{P}$.

The A-optimality error (\ref{Eq:optimisation_al_error}) was solved using an optimisation algorithm, adapted from~\citet{wagenmaker_active_nodate}. For this, the input covariance matrix $\mathbf{\Sigma}_U$ was split between covariance matrix of the $k$ previous stimulations $\mathbf{\Sigma}_{U_{History}} = \mathbf{U}\mathbf{U}^\top  \frac{1}{k}$ and the next stimulation that we want to optimise: $\mathbf{\Sigma}_{U_{Next}} = \mathbf{u}\mathbf{u}^\top$ , i.e. $\mathbf{\Sigma}_U=\mathbf{\Sigma}_{U_{History}}+\mathbf{\Sigma}_{U_{Next}}$. Gradient descent on the loss of the trace of the inverse of the projected covariance matrix ($\text{Tr}\left( (\mathbf{P} \boldsymbol{\Sigma}_U \mathbf{P}^\top)^{-1} \right)$) was then used to optimise $\mathbf{\Sigma}_{U_{Next}}$.A maximum of $10 \times I = 1000$ optimisation steps were carried out, with early stopping if the loss did not improve for the last 10 steps.

At each stimulation point, the `Balanced' method chose randomly among the top $d=3$ left eigenvectors of the currently fitted recurrent system matrix $\hat{\mathbf{W}}$ . Similarly,  two ground truth (also referred to as `GT' or oracle) methods were used which either stimulated uniformly along the top $d=3$ left eigenvectors of the known ground truth recurrent matrix $\mathbf{W}$ (`Balanced GT EVD') or top $d=3$ right singular vectors of the known ground truth recurrent matrix $\mathbf{W}$ (`Balanced GT SVD').  Lastly, the `random’ baseline chose one of the N = 100 neurons randomly to stimulate.

To ensure the inputs are comparable across methods, each stimulation was restricted to be positive and has a maximum L2-norm of $\epsilon=2$ (this budget was chosen through hyperparameter search, aiming for a stimulation strength that is as weak as possible while eliciting a strong enough signal-to-noise ratio). This was accomplished by applying ReLu and then rescaling the chosen input $\mathbf{u}$ to this maximum norm. For the A-optimality methods, this was done after each optimisation step. We also forced inputs to be real, to account for potential small imaginary values arising from EVD. The only exception is the `Balanced GT' (Balanced with known ground-truth eigenvectors) method, where inputs were not required to be positive (but within budget and real).

For all of the active-learning methods and the balanced method an $\epsilon$-greedy approach was used, where the methods stimulation choice was used 80\% of the time, and otherwise a random neuron was stimulated with strength $\epsilon$ (random stimulation).

\subsection{System evaluation}
\label{app:system_eval}
To measure the system identification error, the fitted system matrices $\hat{\mathbf{W}}$, $\hat{\mathbf{B}}$, and $\hat{\mathbf{H}}$ were compared to the discretised ground truth matrices which were projected into the observation space (since the matrices are fitted at the observation level) $\mathbf{W}_{gt}$, $\mathbf{B}_{gt}$, and $\mathbf{H}_{gt}$. For the recurrent and input matrices this is simply: $\mathbf{W}_{gt}=\mathbf{C}_d\mathbf{W}_d\mathbf{C}_d^\top$ and $\mathbf{B}_{gt} = \mathbf{C}_d\mathbf{B}_d = \mathbf{I}$. $\mathbf{H}_{gt}$ was determined by rolling out $\mathbf{W}_{gt}$ and $\mathbf{B}_{gt}$ over the $T_{trial}=300$ time steps for each stimulation trial:
$$\mathbf{H}_{gt}= \sum_{t=0}^{T_{trial}-1} \mathbf{W}_{gt}^{t}\mathbf{B}_{gt}$$
The error in the estimated matrices was then computed using the parameter error: $\frac{\| \mathbf{W} - \hat{\mathbf{W}} \|}{\| \mathbf{W} \|}$. Their non-normality was measured using $\frac{\|\mathbf{W} \mathbf{W}^\top - \mathbf{W}^\top\mathbf{W}\|}{\|\mathbf{W} \mathbf{W}^\top\|}$.

For evaluating the activity prediction ability of the identified system a full and manifold test activity set was created for each system (seed), which was used across all stimulation time points and methods. To accomplish this a set of $K_{test}=10$ test inputs $u_{test}$ were created, pointing in random direction in the full neural space with their L2-norm meeting the budget requirement of $\epsilon=2$. These $K_{test}=10$ stimulation trials were run on both the discretised latent (\ref{Eq:latent_lds})  and full system (\ref{Eq:full_lds}), projected into the observation space, giving us the manifold $\mathbf{Y}_{manifold}\in \mathbb{R}^{K_{test}T_{trial} \times I}$ and full observed activity $\mathbf{Y}_{full}\in \mathbb{R}^{K_{test}T_{trial} \times I}$, respectively. Note, to project the latent activity $\mathbf{z}$ directly into the observation space we defined the latent observation matrix $\tilde{\mathbf{C}}=\mathbf{C}\mathbf{J}^\top$, resulting in $\mathbf{y}_{manifold}(t) = \tilde{\mathbf{C}}\mathbf{z(t)}  + \tilde{\mathbf{v}}(t)$.

To evaluate the full system, the initial activity $\mathbf{y}_{full}(0)\in \mathbb{R}^{I}$ and the stimulation inputs $u_{test}(t_{test})$ at $t_{test}=[0,T_{trial}, \dots, (K_{test}-1)T_{trial}]$ were provided to the fitted system and used to roll-out the activity prediction for the whole $K_{test} \times T_{trial}=3000$ time points. Latent activity was predicted the same way, but with $\hat{\mathbf{W}}$ and $\hat{\mathbf{B}}$ projected onto the top $r=3$ eigenvectors of $\hat{\mathbf{W}}$ to only account for manifold activity, i.e. $\hat{\mathbf{W}}_{p} = \mathbf{Q}_r[:,:r] \mathbf{E}[:r,:r] \mathbf{Q}_l[:r,:]$ and $\hat{\mathbf{B}}_p = \mathbf{Q}_r[:,:r] \mathbf{Q}_l[:r,:]\hat{\mathbf{B}}$ . Accuracy of the latent and main activity prediction were evaluated using the Mean Squared Error (MSE) across all neurons and time points.

To investigate whether the correct eigenvectors and eigenvalues were obtained, the estimated modes were matched to the ground truth based on their eigenvalues. For this we first reconstructed the estimated continuous eigenvalues $\hat{\lambda}_c$ from the fitted discrete eigenvalues $\hat{\lambda}_d$ (diagonal entries of $\hat{\mathbf{E}}_d$ from the EVD of $\hat{\mathbf{A}}_d=\mathbf{Q}_r\hat{\mathbf{E}}_d\mathbf{Q}_l$):

\[
  \hat{\lambda}_c = \frac{\tau}{dt}\hat{\lambda}_d+1-\frac{\tau}{dt},
\]

based on approximating the continuous system as $\mathbf{\hat{A}_c} \approx \frac{\tau}{dt}(\mathbf{\hat{A}_d}-\mathbf{I})+\mathbf{I}$. Note that the difference between estimating $\hat{\lambda}_c$ using this method as opposed to using the exact expression $\mathbf{\hat{A}_c} = \frac{\tau}{dt}\log{(\mathbf{\hat{A}_d})}+\mathbf{I}$ is in the $10^{-3}$ and thus does not affect the ranking of the compared AL methods.
The mode with the third largest estimated eigenvalue was then matched to the first ground truth eigenvector, the mode with the second largest eigenvalue to the second eigenvector, and the one with the largest eigenvalue to the third eigenvector. Error in eigenvalue estimation was quantified as the absolute difference between the ground truth eigenvalues $\lambda$ and the estimated continuous eigenvalue $\hat{\lambda}_d$. Alignment of the matched eigenvector was then assessed using the absolute value of the Cosine similarity: $cos(\alpha) = \mathbf{Q}_r\hat{\mathbf{Q}}_r^\top$. We also used the Cosine similarity to compute the angle in degrees between the first and second estimated eigenvectors ($\hat{\alpha_1}$) and the first and third estimated eigenvector ($\hat{\alpha_2}$).

All Figures show the mean performance across five seeds, with the shaded area being the std.

\subsection{Restricting observability}
\label{app:partial_obs}
For all simulations besides in Appendix~\ref{sec:partialobservation}, the observation $\mathbf{C} \in \mathbb{R}^{I \times N} $ and input $\mathbf{B} \in \mathbb{R}^{N \times I}$ matrices were set to the identity (full observation/control). To generate the partial observations in Appendix~\ref{sec:partialobservation}, $\mathbf{C}$ was defined in two different ways, to create either the readout percentage or the readout angle.

Firstly, to readout $I/N\%$ of the neurons, $I$ neurons were randomly selected among the $N=100$. Each row of $\mathbf{C}$ was then set to each of these $I$ selected neurons, making it a semi-orthonormal matrix only observing a subset of the neurons.

For the second observation condition, the goal was to create an observation matrix which rows were at most angle $\alpha_{obs}$ to one of the eigenvectors (here we used the second right eigenvector $q_{R}^2$). To accomplish this the first row / readout node of $\mathbf{C}$ was chosen to be a unit vector $\mathbf{w}$ at angle $\alpha_{obs}$ to $\mathbf{q}_{R}^2$. All other nodes were then chosen to be in the null-space of $[\mathbf{w},\mathbf{q}_{R}^2]^\top$. Hence, using this method up to $I=N-1=99$ readout vectors were defined (maximum number possible), with $\mathbf{C}$ still being semi-orthonormal. Three different set-ups with $\alpha_{obs} \in [0^\circ,45^\circ,90^\circ] $ were created this way.

We assume that for typical experimental setups recording and stimulation involves the same independent combination of neural groups.
The input matrix was defined to be the transpose of the readout : $\mathbf{B} = \mathbf{C}^\top$. Since the system is evaluated on the observation level the ground truth input matrix for evaluation $\mathbf{B}_d =\mathbf{C}\mathbf{C}^\top=\mathbf{I}$ was still the identity.
The fitted recurrent matrix is already aligned to the observation space, i.e. $\hat{\mathbf{W}} \approx \mathbf{C}\mathbf{W}\mathbf{B}$, with the input matrix being trivial $\hat{\mathbf{B}} \approx \mathbf{C}\mathbf{B}=\mathbf{I}$.

\section{Learning dynamics (eigenvalues) takes more stimulations than identifying manifold geometry (eigenvectors)}
\label{apd_eigvalsvsvec}

\begin{figure}[h]
  \centering
  \includegraphics[width=.8\linewidth]{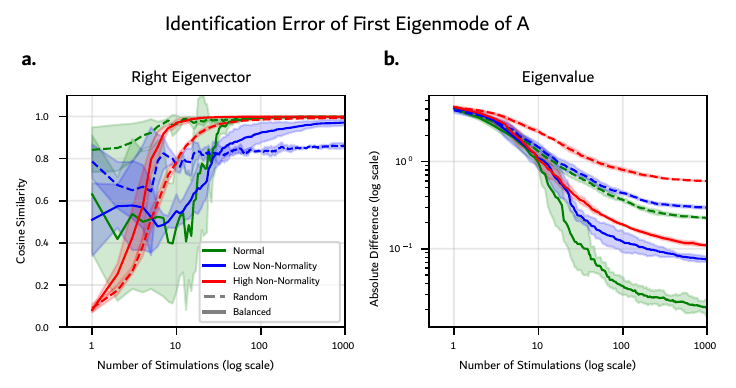}
  \caption{Error in predicting dynamics parameters is not driven by the eigenvector (\textbf{a}) misalignment, but by misidentification of the eigenvalues (\textbf{b}). \textbf{a}. Alignment of the first estimated right eigenvector to the ground truth wrt the number of stimulations for different levels of non-normality and stimulation methods. \textbf{b}. Convergence of the estimated first eigenvalue to the true value.}
  \label{fig: fig3_supplemenatary_evd}
\end{figure}

To better understand the differences in the estimated recurrent matrices for the different non-normality levels and AL methods, the accuracy in identifying eigenvectors and eigenvalues over time was investigated (see Fig.~\ref{fig: fig3_supplemenatary_evd} for the Balanced and random method). For both low and high non-normality, the first eigenvector of $\mathbf{A}$ can be identified independent of the AL method (see Appendix~\ref{app:all_hybrid_methods} for all other strategies and eigenvectors). However, under medium non-normality, random stimulation fails to identify the eigenvector. This is due to the eigenvectors having around a $45^\circ$ angle between them and thus part of the activity of the first eigenvector was explained by the third eigenvector and the other part by a component orthogonal to the third eigenvector, i.e. the estimated first eigenvector (see Appendix~\ref{app:all_hybrid_methods}).
This demonstrates that while the principal directions of variability are unambiguous in normal and strongly non-normal systems, due to activity spreading across orthogonal modes (normal) or collapsing along a single dominant axis (extremely non-normal), systems with low non-normality present a challenge.
However, attributing it to the wrong basis leads to a misidentification of the dynamics along these modes, as can be seen in Fig.~\ref{fig: fig3_supplemenatary_evd}b. Indeed, even for the normal and highly non-normal case, where the eigenvectors are strongly aligned with ground truth, most AL methods are unable to identify the correct eigenvalue of the system, except for our Balanced method.

\section{Under- and over-estimation of non-normal dynamics under partial observation}
\label{sec:partialobservation}

The results above considered full access to observe and stimulate every neuron.
As this is infeasible in experimental settings, it is important to consider how partial observation impacts our system estimation. A common but simplistic approach to model this is random subsampling of individual neurons (10\% to 99\%). However, this `toy model' assumes there is no alignment between individual neurons and the manifold, and, therefore, obscures the fundamental geometric insight: what matters is not the raw number of recorded neurons, but how well the observation space captures the underlying dynamics manifold and whether any critical eigenmodes are missed.

To explicitly investigate this geometric alignment, we designed a second partial observation model that limits the maximal alignment between the system's eigenvectors and the readout nodes (rows in $\mathbf{C}$).
We considered a case where only one basis direction in the neural space is left out, but the observed ones can be either fully aligned ($0^\circ$), partially aligned ($45^\circ$), or orthogonal ($90^\circ$) to the \textbf{second} right eigenvector (see Appendix~\ref{app:partial_obs}).
Missing the second eigenmode, which corresponds to the middle link of the feedforward chain in Fig.~\ref{fig: fig1_setup}c.iii, can theoretically lead to misidentifying a non-normal system as normal according to~\citet{qian2024partial}.

\begin{figure}[h]
  \centering
  \includegraphics[width=.8\linewidth]{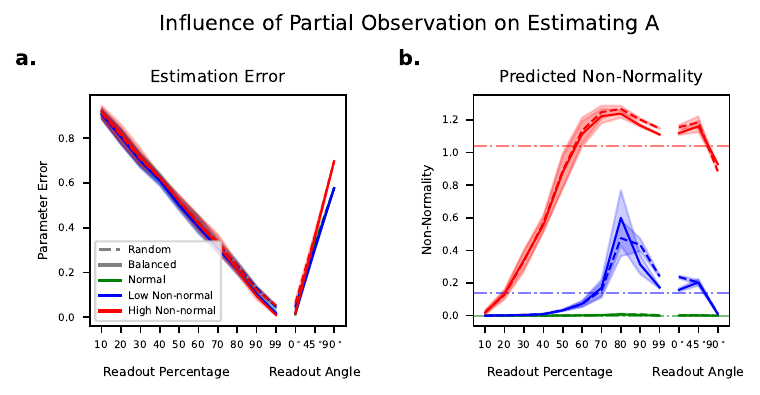}
  \caption{Dynamics parameter estimation error decreases with increased observability (\textbf{a}), while estimated non-normality shifts from underestimation ($<$50\% observability) to overestimation ($>$50\%), before converging to the true value only under full observability (\textbf{b}).(\textbf{a}). Final prediction error of the latent recurrent matrix $\mathbf{A}$ for different non-normality levels, given varying percentage of observed neurons (left: from 10\% to 99\%) or varying maximum angle to the ground truth second eigenvector (right: from $0^\circ$ to  $90^\circ$). (\textbf{b}). Non-normality level of the estimated matrices $\hat{\mathbf{A}}$ given the partial observation conditions, showing that under low observation non-normality is underestimated, but then gets inflated for medium observations. Horizontal lines show ground truth non-normality for normal (green), low non-normal (blue), and high non-normal (red) systems.}
  \label{fig: fig5_partialobservation}
\end{figure}

Firstly,  considering partial observation (left part of the x-axis in Fig.~\ref{fig: fig5_partialobservation}), the final error of the estimated latent matrix reduces approximately linearly with increased observation, independent of non-normality strength. However, estimated non-normality did not follow this trend, with low observation underestimating non-normality but medium observation overestimating it. This can be understood when considering the readout angle for the mixed observation condition (right side of the x-axes). When we can observe all eigenmodes ($0^\circ$), the non-normality of the system can be nearly correctly estimated. However, if at least one of the eigenmodes is only partially observable ($45^\circ$) the non-normality level is overestimated, while completely missing one ($90^\circ$) leads to an underestimation (in agreement with~\citet{qian2024partial}). Therefore, if fewer neurons are observed it is likely at least one eigenmode is not observed at all. With increased access, at least parts of each eigenmode can be observed, leading to over-estimation, and only under close to full observation do we then get enough information about the full system to correctly estimate the non-normality.

\section{Approximating non-linear systems using linear AL approaches}
\label{app:non-linear}

While the linear ground truth assumptions allowed us to gain analytical insights into the effectiveness of the compared methods, neural dynamics are inherently non-linear. Thus, to understand whether the linear AL approaches are still effective in predicting neural activity even under a model mismatch, we simulated a non-linear ground truth system. As for the linear simulations, we first define the latent dynamics matrix $\mathbf{A}$ for the three different non-normality levels. The linear model was then adapted to have a non-linear transformation of the latent activity:

\begin{equation}
  \tau \frac{d\mathbf{z}}{dt} = \mathbf{A} \operatorname{ReLU}(\mathbf{z}(t)) - \mathbf{z}(t) + \tilde{\mathbf{B}}\mathbf{u}(t)  + \tilde{\mathbf{w}}(t),
  \label{Eq:latent_nlds}
\end{equation}
where $\operatorname{ReLU}(z)=max(0,z)$ denotes the element-wise Rectified Linear unit activation function.

Using the same projection as before with $\mathbf{z}=\mathbf{J} \mathbf{x}$, this then leads to neural dynamics of  the form :

\begin{equation}
  \tau \frac{d\mathbf{x}}{dt} = \mathbf{J}^\top  \mathbf{W} \operatorname{ReLU}(\mathbf{J}\mathbf{x}(t)) - \mathbf{x}(t) +  \mathbf{B}\mathbf{u}(t) + \mathbf{w}(t),
  \label{Eq:full_nlds}
\end{equation}

where neural dynamics are restricted to be along the manifold while having both on and off manifold leak terms and inputs.

The corresponding discretised versions (using Euler method) for activity simulation were then respectively:

\begin{equation}
  \tau \mathbf{z}(t+1) = \frac{dt}{\tau}\mathbf{A} \operatorname{ReLU}(\mathbf{z}(t)) + (1-\frac{dt}{\tau})  \mathbf{z}(t) + \frac{dt}{\tau}\tilde{\mathbf{B}}\mathbf{u}(t)  + \tilde{\mathbf{w}}(t)
  \label{Eq:latent_nlds_disc}
\end{equation}

and

\begin{equation}
  \tau \mathbf{x}(t+1) = \frac{dt}{\tau}\mathbf{J}^\top  \mathbf{W} \operatorname{ReLU}(\mathbf{J}\mathbf{x}(t)) -(\mathbf{z}(t)) + (1-\frac{dt}{\tau})   \mathbf{x}(t) +  \frac{dt}{\tau}\mathbf{B}\mathbf{u}(t) + \mathbf{w}(t),
  \label{Eq:full_nlds_disc}
\end{equation}

with the readouts  $\mathbf{y}_{manifold}(t) = \tilde{\mathbf{C}}\mathbf{z}(t)  + \tilde{\mathbf{v}}(t)$ and  $\mathbf{y}_{full}(t) = \mathbf{C}\mathbf{x(t)}  + \mathbf{v}(t)$ . All parameters were set to the same values as for the linear case, assuming full observability (see Appendix~\ref{app:comp_experiments}). System identification, stimulation selection, and evaluation were carried out as before, fitting a linear dynamical system to the simulated activity. The only difference is that regularisation was set to be stronger with $\lambda_W=\lambda_H=100$ and $K=2500$ stimulation trials being simulated. Since there are no ground truth matrices to compare to, we only evaluated the activity prediction of the stimulation approaches.

\begin{figure}[h]
  \centering
  \includegraphics[width=.8\linewidth]{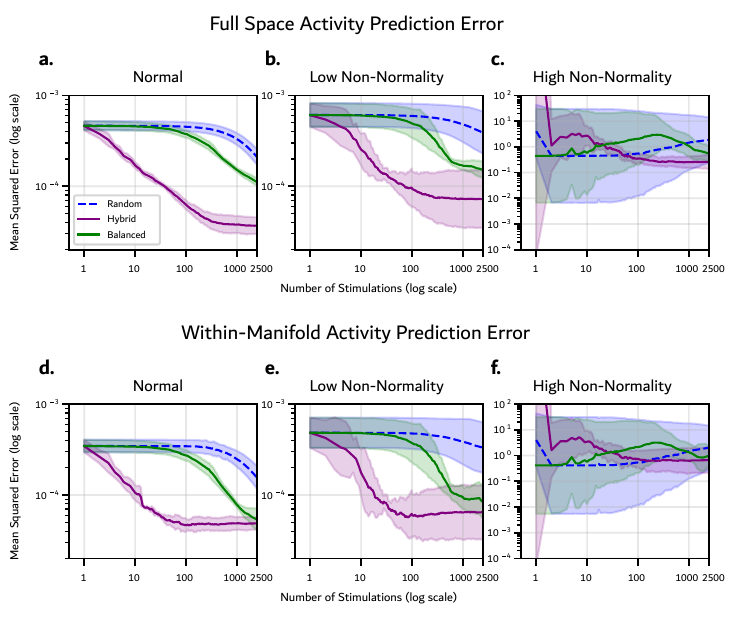}
  \caption{Active Learning methods lead to more precise prediction of non-linear activity than random stimulations. Estimation error of full system (top row) and within-manifold activity of a non-linear system with normal (\textbf{a.}, \textbf{d.}), low non-normal (\textbf{b.}, \textbf{e.}), and high non-normal (\textbf{c.}, \textbf{f.}) recurrent dynamics, approximated by a linear dynamical system.}
  \label{fig: fig_non-linearsystem}
\end{figure}

We compared the effectiveness of the random baseline to the two best performing AL approaches from the linear results: Hybrid using SVD as a basis and Balanced. As can be seen in Fig~\ref{fig: fig_non-linearsystem}, for normal and low non-normal systems both the full and on-manifold non-linear activity can be approximated using a linear dynamical system. While it takes more sample points for the systems to converge, the AL methods are still more efficient than selecting random perturbations. Notably, the Hybrid approach is notably quicker in system identification, with the Balanced method only catching up after more than 2000 sample points. This indicates that while prioritizing stimulations along the estimated left eigenvectors of the recurrent matrix appears to be a powerful heuristic for the linear case, this advantage might gets lost for non-linear system since the activation term removes the relationship between the eigenvectors of $\mathbf{A}$ and $\mathbf{W}$ for negative valued latent states ($\mathbf{Q}_l^{(\mathbf{W})}\neq \mathbf{Q}_l^{(\mathbf{A})}\mathbf{J} $ for $\mathbf{z} < 0$).
Lastly, for the highly non-normal non-linear dynamics, there appears to be no linear dynamical system that can approximate it, given that none of the stimulation approaches are able to converge to a low activity estimation error.

\section{Performance of other hybrid methods and baselines}
\label{app:all_hybrid_methods}

\begin{figure}[h!]
  \centering
  \includegraphics[width=.8\linewidth]{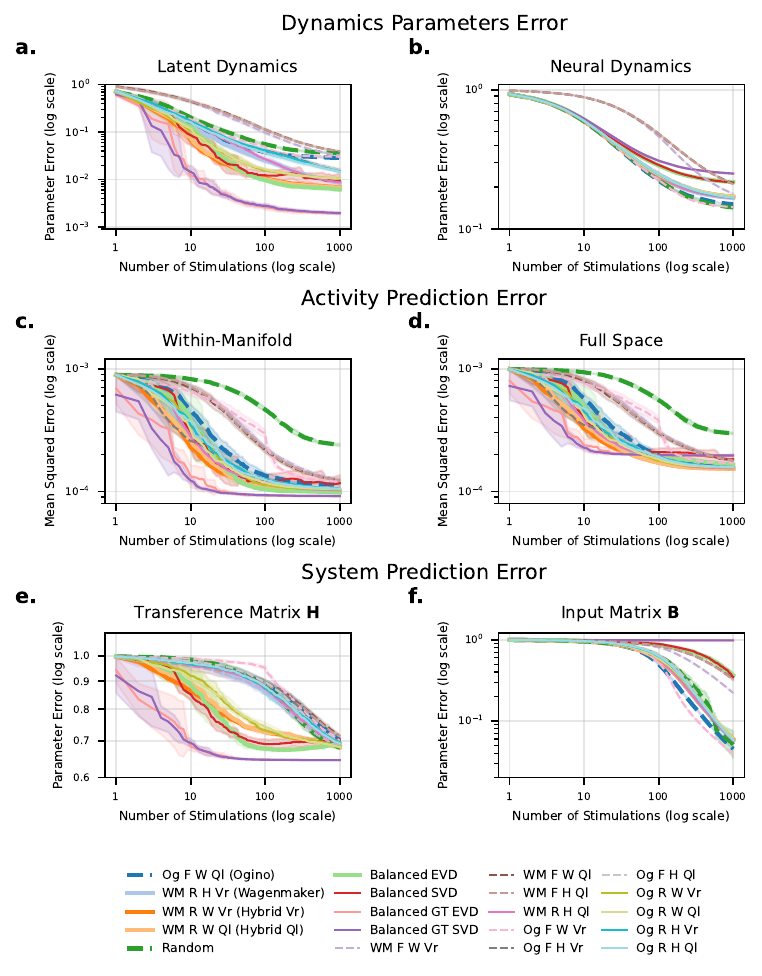}
  \caption{System identification errors for all 16 combinations of A-optimality, random, ground truth, and balanced stimulation for a normal system, extended version of Fig.~\ref{fig: fig2_manifold}.}
  \label{fig: apd_fullobs_normal_DynAndActv}
\end{figure}

\begin{figure}[h!]
  \centering
  \includegraphics[width=.8\linewidth]{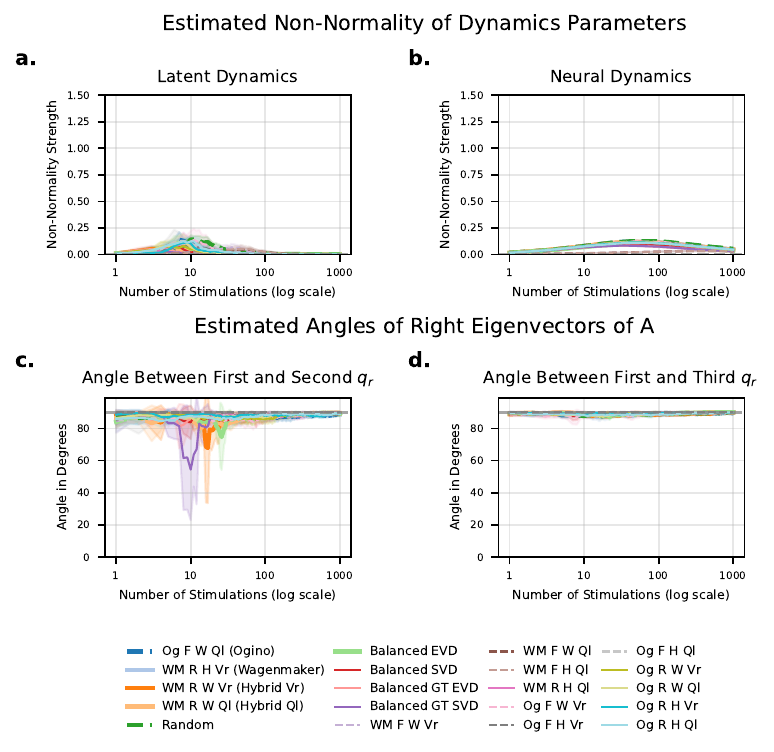}
  \caption{Estimated non-normality of the recurrent matrices for all 16 combinations of A-optimality, random, ground truth, and balanced stimulation for a normal system (grey horizontal dashed line is ground truth).}
  \label{fig: apd_fullobs_normal_NNAndAngle}
\end{figure}

\begin{figure}[h!]
  \centering
  \includegraphics[width=.8\linewidth]{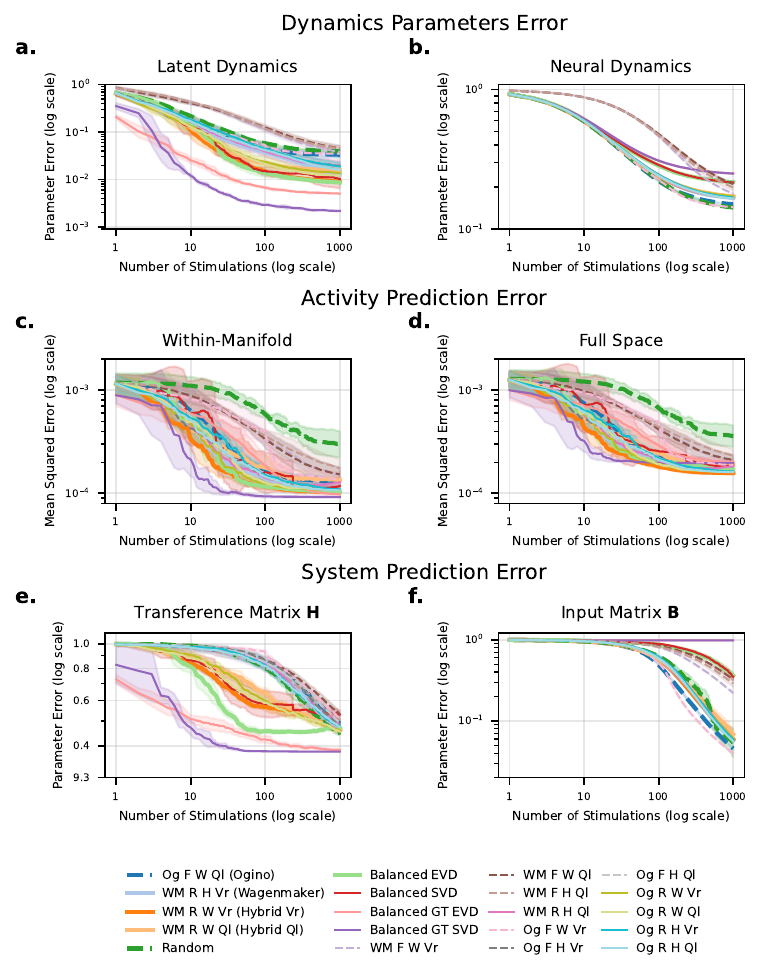}
  \caption{System identification errors for all 16 combinations of A-optimality, random, ground truth, and balanced stimulation for a low non-normal system.}
  \label{fig: apd_fullobs_lownnonnormal_DynAndActv}
\end{figure}

\begin{figure}[h!]
  \centering
  \includegraphics[width=.8\linewidth]{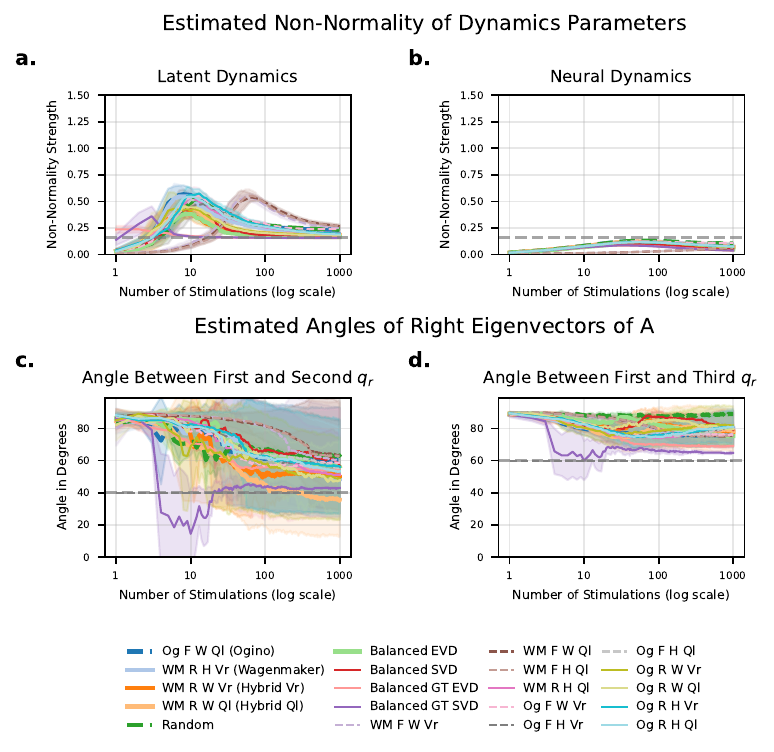}
  \caption{Estimated non-normality of the recurrent matrices for all 16 combinations of A-optimality, random, ground truth, and balanced stimulation for a low non-normal system (grey horizontal dashed line is ground truth).}
  \label{fig: apd_fullobs_lownonnormal_NNAndAngle}
\end{figure}

\begin{figure}[h!]
  \centering
  \includegraphics[width=.8\linewidth]{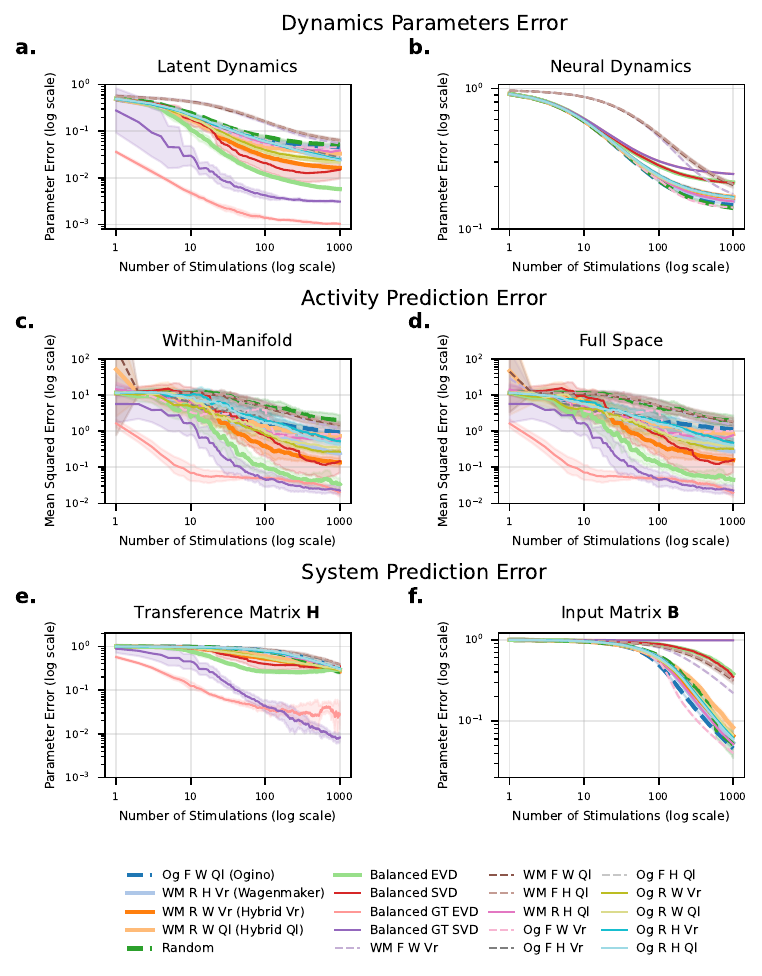}
  \caption{System identification errors for all 16 combinations of A-optimality, random, ground truth, and balanced stimulation for a highly non-normal system.}
  \label{fig: apd_fullobs_highnonnormal_DynAndActv}
\end{figure}

\begin{figure}[h!]
  \centering
  \includegraphics[width=.8\linewidth]{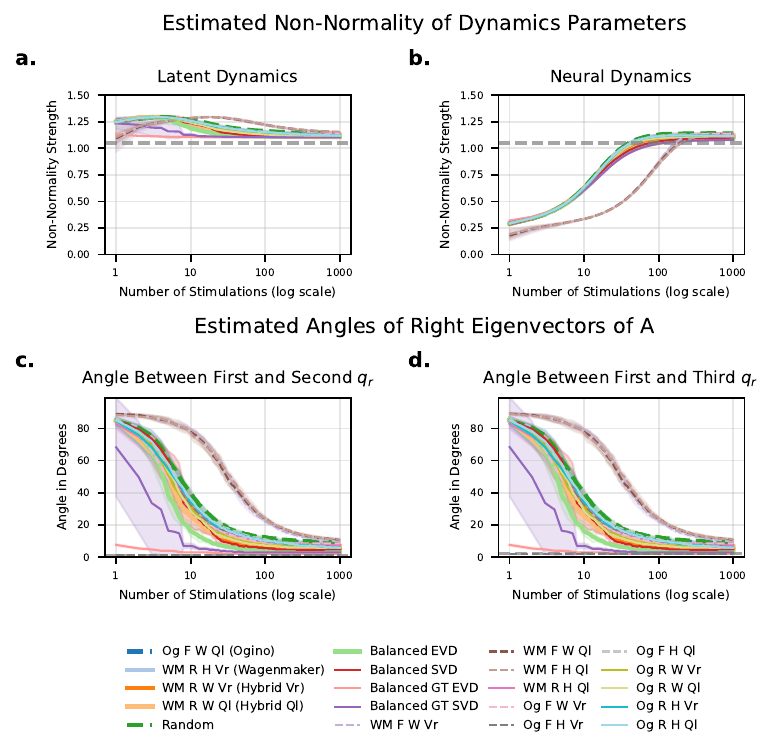}
  \caption{Estimated non-normality of the recurrent matrices for all 16 combinations of A-optimality, random,ground truth, and balanced stimulation for a highly non-normal system (grey horizontal dashed line is ground truth).}
  \label{fig: apd_fullobs_highnonnormal_NNAndAngle}
\end{figure}

\begin{figure}[h!]
  \centering
  \includegraphics[width=.8\linewidth]{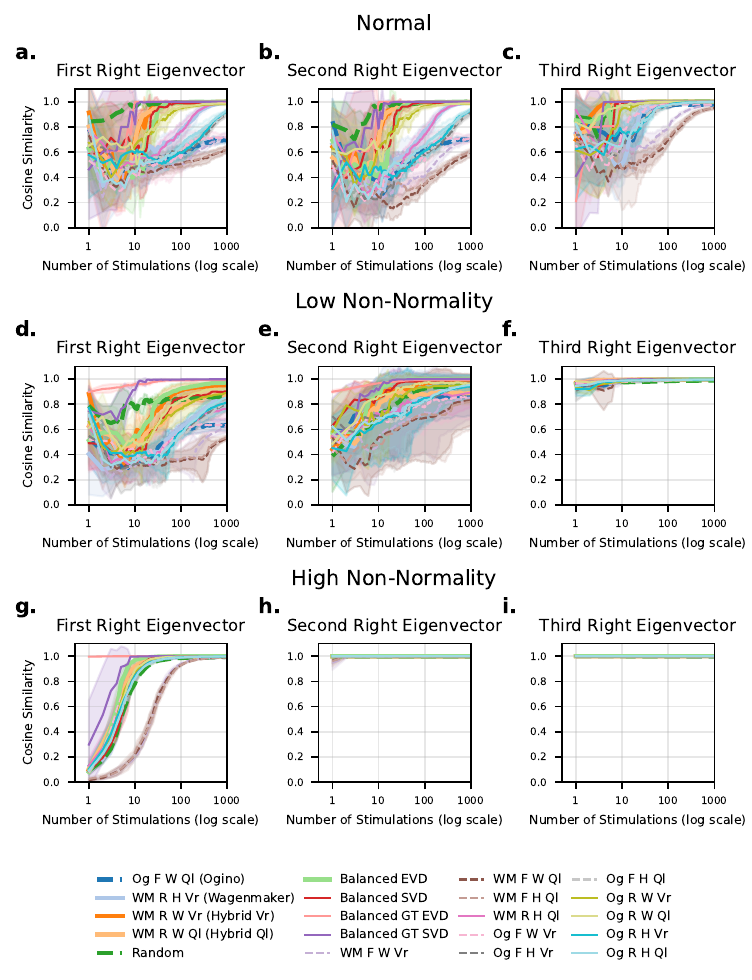}
  \caption{Top three eigenvector alignment for all 16 combinations of A-optimality, random, ground truth, and balanced stimulation for all non-normality levels.}
  \label{fig: apd_fullobs_allnonnormal_Qr}
\end{figure}

\begin{figure}[h!]
  \centering
  \includegraphics[width=.8\linewidth]{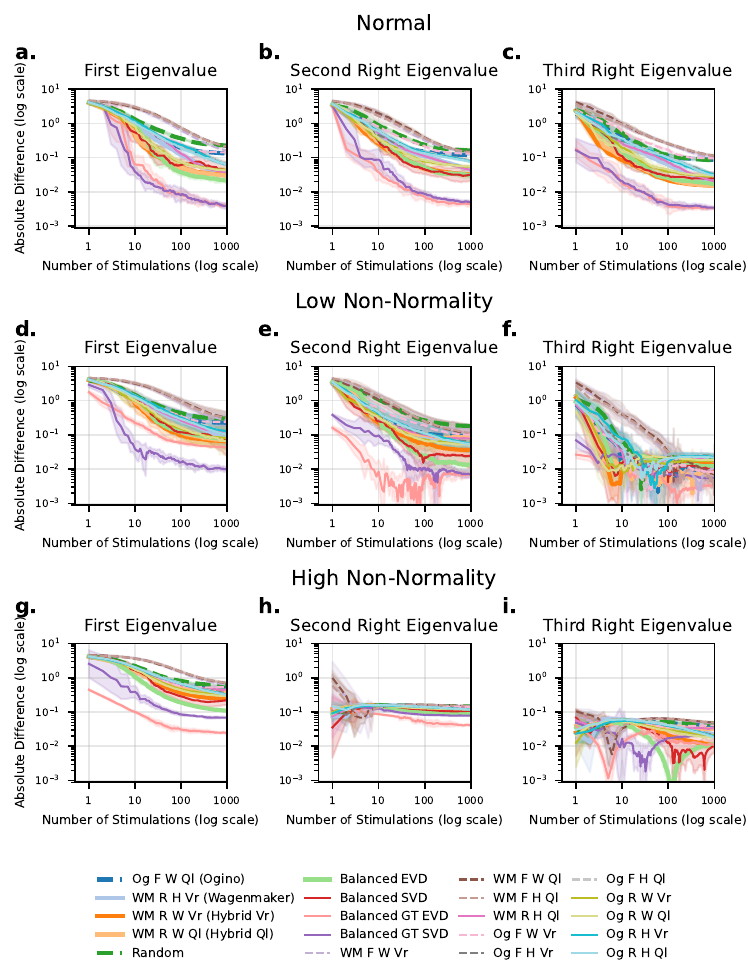}
  \caption{Estimation error of the top three eigenvalues for all 16 combinations of A-optimality, random, ground truth, and balanced stimulation for all non-normality levels.}
  \label{fig: apd_fullobs_allnonnormal_E}
\end{figure}

\begin{figure}[h!]
  \centering
  \includegraphics[width=.8\linewidth]{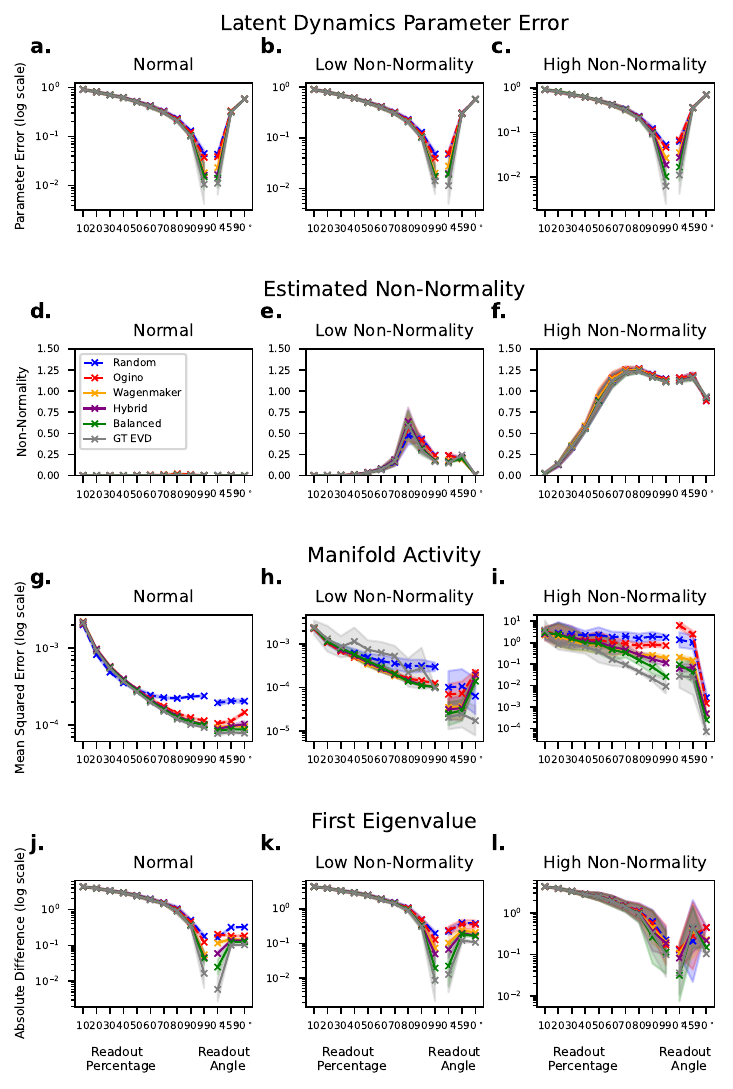}
  \caption{Final latent system identification error wrt to partial observation strength for all non-normality levels for the main considered methods and baseline.}
  \label{fig: apd_partialobs}
\end{figure}

This appendix provides a comprehensive evaluation of all 16 active learning method combinations. The following figures detail the system identification errors (Figures~\ref{fig: apd_fullobs_normal_DynAndActv}, \ref{fig: apd_fullobs_lownnonnormal_DynAndActv}, and \ref{fig: apd_fullobs_highnonnormal_DynAndActv}) and estimated non-normality (Figures~\ref{fig: apd_fullobs_normal_NNAndAngle}, \ref{fig: apd_fullobs_lownonnormal_NNAndAngle}, and \ref{fig: apd_fullobs_highnonnormal_NNAndAngle}) across normal, low, and high non-normality levels, respectively. Additionally, we provide the extended eigenspace alignments (Figures~\ref{fig: apd_fullobs_allnonnormal_Qr} and~\ref{fig: apd_fullobs_allnonnormal_E}) and evaluations under partial observability constraints (Figure~\ref{fig: apd_partialobs}).


\end{document}